\RequirePackage{fix-cm}
\documentclass[twocolumn]{svjour3}          % twocolumn
\smartqed  % flush right qed marks, e.g. at end of proof
\usepackage{wrapfig}
\usepackage{graphicx}
\usepackage{threeparttable}
\usepackage{tablefootnote}
\usepackage{lineno}
\usepackage{multirow}% http://ctan.org/pkg/multirow
\usepackage{hhline}
\usepackage{xcolor}
\usepackage{float}
\usepackage[autostyle]{csquotes}

\journalname{- Journal}
\begin{document}
\title{Leveraging Big Data Analytics in Healthcare Enhancement: Trends, Challenges and Opportunities}

\author{Arshia Rehman        \and
        Saeeda Naz \and
        Imran Razzak %\and Ibrahim A Hameed%etc.
}
\institute{A. Rehman \at Computer Science Department, Govt. Girls Postgraduate College No.1, Abbottabad, KPK, Pakistan \\
\email{arshiar29@gmail.com}
    \and
           S. Naz \at  Computer Science Department, Govt. Girls Postgraduate College No.1, Abbottabad, KPK, Pakistan \\
              \email{saeedanaz292@gmail.com}
    \and
           I. Razzak \at
           Deakin University, Geelong, Australia \\
 \email{imran.razzak@ieee.org}
}
\maketitle
\begin{abstract}

 Clinicians decisions are becoming more and more evidence-based meaning in no other field the big data analytics so promising as in healthcare. Due to the sheer size and availability of healthcare data, big data analytics has revolutionized this industry and promises us a world of opportunities. It promises us the power of early detection, prediction, prevention and helps us to improve the quality of life. Researchers and clinicians are working to inhibit big data from having a positive impact on health in the future. Different tools and techniques are being used to analyze, process, accumulate, assimilate and manage large amount of healthcare data either in structured or unstructured form. In this paper, we would like to address the need of big data analytics in healthcare: why and how can it help to improve life?. We present the emerging landscape of big data and analytical techniques in the five sub-disciplines of healthcare i.e.medical image analysis and imaging informatics, bioinformatics, clinical informatics, public health informatics and medical signal analytics. We presents different architectures, advantages and repositories of each discipline that draws an integrated depiction of how distinct healthcare activities are accomplished in  the pipeline to facilitate individual patients from multiple perspectives. Finally the paper ends with the notable applications and challenges in adoption of big data analytics in healthcare.

%  \textcolor{blue}{Big data analytics  in healthcare industry have  revolutionized  its  remarkable  diversity in Information Technology, Social Media, Internet of Things, Government, international development, manufacturing, media,  pharmaceutical and data-driven researches. One of the pharmaceutical research domain in which big data is evolved over the few years is healthcare. Huge amount of clinical, bioinfromatics, imaging and biological data have been generated, analyzed, processed and collected on daily bases at continuous speed and scale. Thus big data analytics provided different tools and techniques to analyze, processed, accumulate, assimilate and manage large amount of healthcare data either in structured or unstructured form. These big data analytical tools and techniques has improved the quality of healthcare, care delivery, patient management and disease exploration. This survey presents the emerging landscape of big data and analytical techniques in the five sub-disciplines of healthcare i.e.Medical Image Analysis and Imaging Informatics, Bioinformatics, Clinical Informatics, Public Health Informatics and Medical Signal Analytics. We presents different architectures, advantages and repositories of each discipline that draws an integrated depiction of how distinct healthcare activities are  accomplished  in  a  pipeline  to  facilitate  individual patients from multiple perspectives. Finally the paper ends with the notable applications and challenges in adoption of big data analytics in healthcare.}

\keywords{Big Data Analytics \and, Medical Image Processing and Imaging Informatics \and Bioinformatics and Genomics \and Clinical informatics \and Public Health informatics \and Medical Signal Analytics}
\end{abstract}
\section{Introduction} \label{sec:introduction}
Due to the sheer size and availability of multidimensional data,  the rate of technological innovation have the huge potential to make a an extra ordinary impact on our daily life in different disciplines especially in healthcare sector. The rapidly growing and exploited data will refer to introduce a new gigantic term known as big data. Uncovering information from such complicated nature of data is often complex process. The development and analysis of tools and methods for analysis of such large quantities of data provides us with an opportunity to make the transition into this new era far easier. Having data-driven, real-time insights accessible to the organization through analytics can be a critical enabler for executing the organization strategies. Big data analytic’s greatest asset is its possibilities and its need to find new ways to provide the services that we are looking for.

Unlike other field, big data analytics is so promising in healthcare sector and received much more attention in the last few years. Clinicians decisions are becoming evidence-based, meaning that they are relying more on large swathes of research and clinical data as opposed to solely their schooling and professional opinion. Big data in terms of healthcare is defined as the name given to larger and complex electronic healthcare datasets that are problematic or almost impossible to manage by employing common traditional methods, tools or software \cite{frost2015drowning,raghupathi2014big,baro2015toward}. Big data in healthcare is generated by healthcare record (such as patients record, disease surveillance, hospital, medicine, health management, doctor, clinical decision support or feedback of patient  \cite{burghard2012big,dembosky2012data,feldman2012big,fernandes2012big}) and clinical data (like imaging, personal, financial record, genetic and pharmaceutical data and Electronic Medical Records (EMR) etc. \cite{vayena2015ethical,naseer2019refining}). The generation and management of these enormous healthcare records is considered to be very complex thus, big data analytics is introduced \cite{wyber2015big,ward2014applications}. With the rise of technological innovation and personalised medicine, big data analytics has the potential to make a huge impact on our life i.e. how it helps to predict, prevent, manage, treat and cure disease. Furthermore, it helps, government agencies, policy maker and hospital to manage resources, improving medical research, planning preventative methods and managing epidemic.

\begin{figure}[hbtp]
\begin{center}
  \includegraphics[width=8cm,height=6cm]{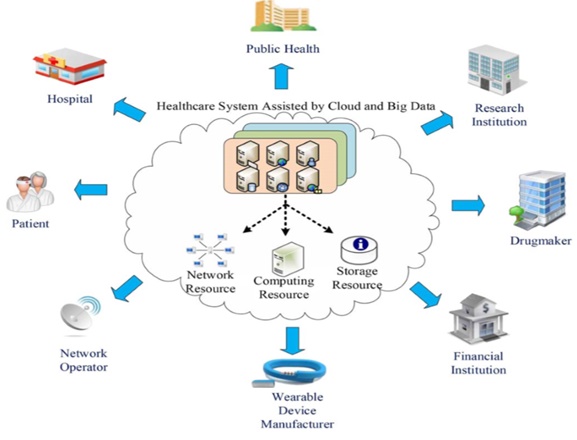}
  \caption{Healthcare Ecosystem assisted By big data and Cloud Computing \cite{manogaran2017survey}}
\label{fig:Ecosystem}  
\end{center}% Give a unique label
\end{figure}

With the advancement in information technology and emergence of digitized computerized systems, hard copy medical data is tend to move towards Electronic Health Records (EHR) and Electronic Medical Records (EMR) systems. These systems generated exponential growth of data \cite{sessler2014big,razzak2019big}.  Health data is not only collected from clinical record, tele-monitoring or medical tests but there are also a larger number of healthcare apps. These apps have tremendous amount of subscriptions.  According to the \textit{Ericsson Mobility Report} of 2018, in Q4 of 2017, there were a total of 7.8 billion mobile subscriptions, with 53 million new subscriptions added during the quarter as the growth of people on this planet subscribe new and valuable data about health and well-being everyday. These apps contain voluminous data due to the world of social media. There are more than two billions people who use internet for the purpose of mailing, downloading, surfing, blogging and entertainment etc. This amount of data also tend to move towards the concept of big data. Fig.~\ref{fig:Ecosystem} depicts the ecosystem of healthcare assisted by big data and cloud computing approaches.

Moving towards the five characteristics of big data in healthcare sector, \textit{Volume} refers to the medical record of personal data, clinical data, radiology images, genetics and population information, resource intensive applications like 3D imaging genomics and biological sequences.  Likewise rapid increase in diseases and medications produce exponential growth of data that is to be stored, manipulate and managed.  For the effective capturing, management and manipulation of data, modern techniques like advances in data management, cloud computing and visualization etc. play a vibrant role for healthcare systems. Volume is rapidly increasing in bio-medical informatics like Proteomics DB \cite{wilhelm2014mass} contains data volume of 5.17 TB covering 92\% of human genes information explained in Swiss-Prot database. Vast amount of volume is produced from medical images like Visible Human Project comprehends female data-sets of 39 GB \cite{ackerman1999visible}. It is estimated that volume of big data in healthcare increased to  35 zeta-bytes by 2020 \cite{gui2016architecture,razzak2018deep}.

\textit{Variety} in healthcare divulges that there is a gigantic amount of healthcare record either it is structured, unstructured or semi structured.  There is a verity of unstructured healthcare record generated daily like patient information, doctor notes, prescriptions, clinical or official medical records, images of MRI, CT and radio films etc. Furthermore, structured and semi structured verity regarding to EMS and EHS comprises actuarial data, electronic apps and automated databases information like physician name, hospital name, treatment reimbursement codes, patient name, address etc., information of electronic billings and accounting and some of the clinical and laboratory instrument reading observations. For the conversion of unstructured data into structured data-sets, data analytics provides different facilities; one of them is natural language processing in health fidelity.

Another important characteristic is \textit{velocity} that can be at rest or motion pace. At rest velocity, healthcare record encompasses doctor or nurse notes, scripts, documentary files, renders record, X-ray films etc. Moreover, medium velocity healthcare data includes blood pressure readings, measurement of daily diabetic glucose by insulin pumps and EKGs etc. However, sometimes high velocity is required, as it become a staple of life or death. This type of data embroils on real time data like monitoring of inside heart, anesthesia and trauma for blood pressure, room operations, detecting infections or diseases like cancer etc. at early stage.

\textit{Value} describes how much data is beneficial or hcare ecosystem. For example raw data like paper prescriptions, official record or patient information is less valuable than diagnostics record, medicines and laboratory instruments reading record. \textit{Veracity} tells the reliability or understandability of healthcare record that explains the capturing of diagnosis, procedures, treatments etc. and to verifying the information of patient, hospital and reimbursement code etc.
Different domains of healthcare and medical care propose in the literature. This review paper discusses five sub-disciplines (i.e., medical image processing and imaging informatics, bioinformatics, clinical informatics, public health informatics, medical signal analytics) that directly or indirectly involve in healthcare and bio-medical \cite{razzak2017microscopic,razzak2015malarial}. Before presenting the literature review, we present the theoretical information of big data and data analytics in Section 2. Different architectures of big data analytics deployed in the domain of healthcare are explaining in Section 3. We also present the advantages of big data to healthcare in Section 4 that give the insights how healthcare can be improved by big data analytics. Then we move towards the literature review for which we have proposed a review methodology for the selection of articles explained in Section 5. Based on the review methodology, the big data in five sub-disciplines of healthcare (i.e., medical image processing and imaging informatics, bioinformatics, clinical informatics, public health informatics, medical signal analytics) comprehensively explain in Section 6. We also summarize our main findings in Section 7. Then, Section 8 presents the notable applications of healthcare analytics based on the main findings. Section 9 discusses the challenges and open research issues. Finally the Section 10  draws conclusion of this paper.

\section{Background of Big Data and Data Analytics}

The concept of big data was introduced in 1990's by Cox and Ellsworth \cite{cox1997application}, when they considered visualization as a Big Data problem. The significant academic references of big data in computer science was first discovered by Weiss and Indurkhya  \cite{weiss1998predictive}. In 2000, Diebold \cite{diebold2003big} introduced big data in statistics/econometrics when they referred to exploited quality information. The concept was enriched by Douglas Laney at Gartner in an unpublished 2001 research \cite{laney20013d}. In short, the term Big data is attributed to Weiss and Indurkhya, Diebold, and Laney.Big data is the name given to the larger and enormous data-sets that are usually complex so that traditional information processing techniques are not enough to deal with them. Mostly the difficulties or challenges regarding to big data are how to capture, store, share and analyze data, how to visualize, update or query information privacy. From the view of Radar   \cite{oreilly}, Big Data deals with the huge amount of data that is not fit into the conventional databases thus alternative way is chosen to extract and process the data from it. According to ZDNet \footnote{Http://www.zdnet.com/blog/virtualization/what-is-big-data/1708} big data involves techniques and procedures for the creation, formation, manipulation and organization of larger data-sets and facilities offering for its storage. Techopedia \footnote{Https://www.techopedia.com/definition/27745/big-data} demarcatdes that unstructured large complex data that is processed by massive parallelism on readily-available hardware because relational database engines are unable to process that data. Literature divulges that big data is larger data sets, enormous growth of data, massive data, unstructured or complex data~\cite{shin2016demystifying,emani2015understandable,chen2014big,groves2013big,eynon2013rise}.

\begin{figure}
\begin{center}
  \includegraphics[width=6cm,height=4cm]{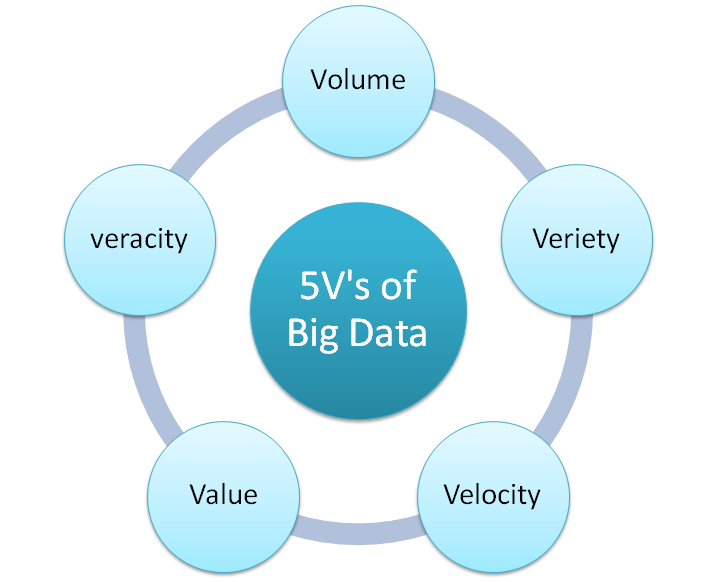}
  \caption{5V's Characteristics of Big Data }
\label{fig:5V}     
\end{center}
\end{figure}

Basically main characteristics of big data are complexity and massive size \cite{porche2014men,berger2014big,bernard2014supporting}. However, big data is deliberated by three characteristics known as 3Vs – volume, variety and velocity \cite{watson2014tutorial,mcafee2012big,russom2011big}. Two additional characteristics are extended to make 5Vs properties of big data as depicted in Fig.~\ref{fig:5V}. These additional characteristics are – value and veracity \cite{emani2015understandable,Saporito,sathi2012big}. Volume leads to the size or quantity of stored and generated data. When the volume of data is large it becomes big data \cite{manogaran2018health,manogaran2017survey}. Variety is the type or nature of data when grouped from several sources. Data is varied in terms of format like CSV, text or Excel format in which data stored in a database. Likewise various forms of data also vary such as video, audio, SMS or PDF data \cite{manogaran2017survey}. This verity is also one of the decisive characteristic of big data. Velocity specifies the speed of data at which it is generated or processed. Value describes how much data is beneficial or valuable. The big data and the value is strongly co-related as storage of raw data is useless and inoperable. Huge data is valuable due to the costs and benefits while collecting and evaluating data \cite{manogaran2017survey}. The term veracity is the quality of data understand-ability. In other words reliability, quality and accuracy of big data depend on the veracity property because it prevents dirty data.

Data analytics is the amalgamation of two words where data refers to raw facts, figures and information and analytics means use of several tools to analyze data although data is small or big. Analytics is a canopy and umbrella term for all data analysis applications \cite{watson2014tutorial}. The big data analytics is the process of analyzing large voluminous data using different strategies. As aforementioned big data is integrated from multiple sources, thus big data analytics is used to explore how to extract valuable and hidden patterns and connections from this integrated data. In other words, big data analytics is simply analysis of data with the intention of extracting information and supporting conclusion making from the inclusive procedure of scrutinizing, modeling, cleansing, and transforming of Big data.

Data analytics can be analyzed by three general methods: descriptive, predictive and prescriptive analytics  \cite{ganjir2016big,bochare2011heterogeneous}. Descriptive analytics deals with the condensation of big data into smaller meaningful information. Predictive analytics is the data reduction analytics that predicts the future analysis by deploying a diversity of machine learning, statistical, modeling and data mining techniques to study latest recent and historical data. Prescriptive analytics is basically the predictive analytics that is used to take action and make the business decision.

Most extensively used approaches for predictive and descriptive analytics on big data are based on either supervised, unsupervised, or hybrid machine learning. An exponential time increase in data has made it difficult to extract valuable information from this data. Despite the strong performance of traditional methods, their predictive power is limited as traditional analysis only deals with primary analysis whereas data analytics deals with secondary analysis.  Data mining involves the digging or mining of data from many dimensions or perspectives through data analysis tools to find previously unknown patterns and associations from data that may be used as valid information \cite{razzak2018robust,naz2017b,razzak2019integrating,naz2016}. Moreover, it makes use of this extracted information to build predictive models. It has been deployed intensively and extensively by many organizations, especially in the healthcare sector.

Data mining is not a magical wand but in fact a big powerful tool that does not discover solutions without guidance. Data mining is convenient for the succeeding purposes:

\begin{itemize}
    \item Exploratory data analysis to examine the data corpus to summarize their main characteristics.
    \item Descriptive modeling to segregating the data into clusters based on their properties. 
    \item Predictive modeling to forecasting information from existing data. 
    \item Discovering pattern to find patterns that occur frequently.
    \item Content retrieval to discover hidden patterns.
\end{itemize}

 Several techniques  deploy for reduction, optimization or regression analysis  etc. for big data. On account of the voluminous amount of big data; its dimensionality is reduced by linear mapping approaches like Principal Component Analysis (PCA)~\cite{holland2008principal}, Singular Value Decomposition (SVD)~\cite{svd2014singular}. Some non linear mapping methods for dimensonality reduction are  Kernel Principal Component Analysis (KPCA) \cite{scholkopf1997kernel}, Sammon’s mapping \cite{sammon1969nonlinear,de1997sammon},Laplacian eigenmaps \cite{belkin2003laplacian}. 

Mathematical optimization is another analytics tool that involve multi-objective and multi-modal
optimization approaches like pareto optimization~\cite{pareto1964cours,horn1994niched}, evolutionary algorithms  \cite{deb2001multi,back1997evolutionary}.
Extracting meaningful information and cluster development and analysis is achieved by various clustering algorithms like Clustering LARge Applications (CLARA) \cite{kaufman2009finding} and Balanced Iterative Reducing using Cluster Hierarchies (BIRCH) \cite{zhang1996birch} etc.
\section{Architectures For Big Data Analytics}
\begin{figure}[hbtp]
\begin{center}
  \includegraphics[width=8cm,height=6.6cm]{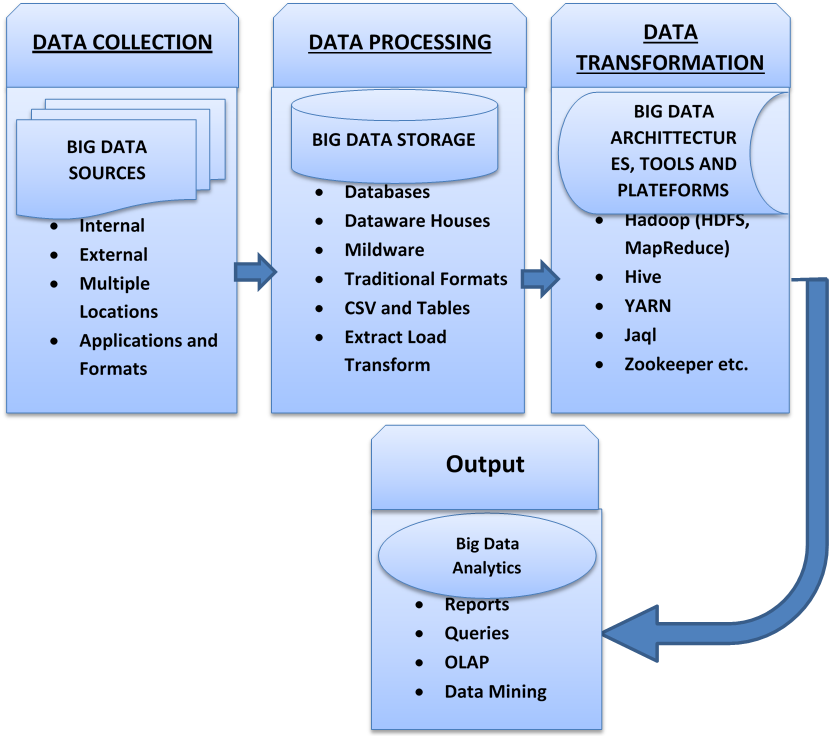}
  \caption{ Conceptual Journey of Data to Information in Big Data Analytics Environment}
\label{fig:BigDataArchitecture}       % Give a unique label
\end{center}
\end{figure}

Our anticipated general framework of big data analytics for healthcare is an abstraction of several conceptual steps that describe the generic functionalities of the domain. The first step in the framework is data collection, in which health and the clinical data is collected from internal or external sources. Verity of data includes Electronic Healthcare Records (EHRs), clinical images and health monitoring devices logs etc. After the collection of data, next step is Data processing in which healthcare data is stored, extract and load in the data ware houses, middle-ware or in traditional formats like CSV, tables etc. Data transformation is the next step in which data is transform, aggregate and loaded in database file systems like Hadoop cloud or in a  Hadoop distributed file systems (HDFS). Analytical phase is used to examine the big data using big data tools and platforms like Hadoop, Mapreduce, Hive, Hbase, Jaql, Avro and several others. Finally the output is generated in the form of reports and queries using data mining and OLAP tools.  The self explanatory general and conceptual architecture are depicted in Fig.~\ref{fig:BigDataArchitecture} and Fig.~\ref{fig:BigDataArchitectures}. 

\begin{figure*}[hbtp]
\begin{center}
  \includegraphics[width=16cm,height=8cm]{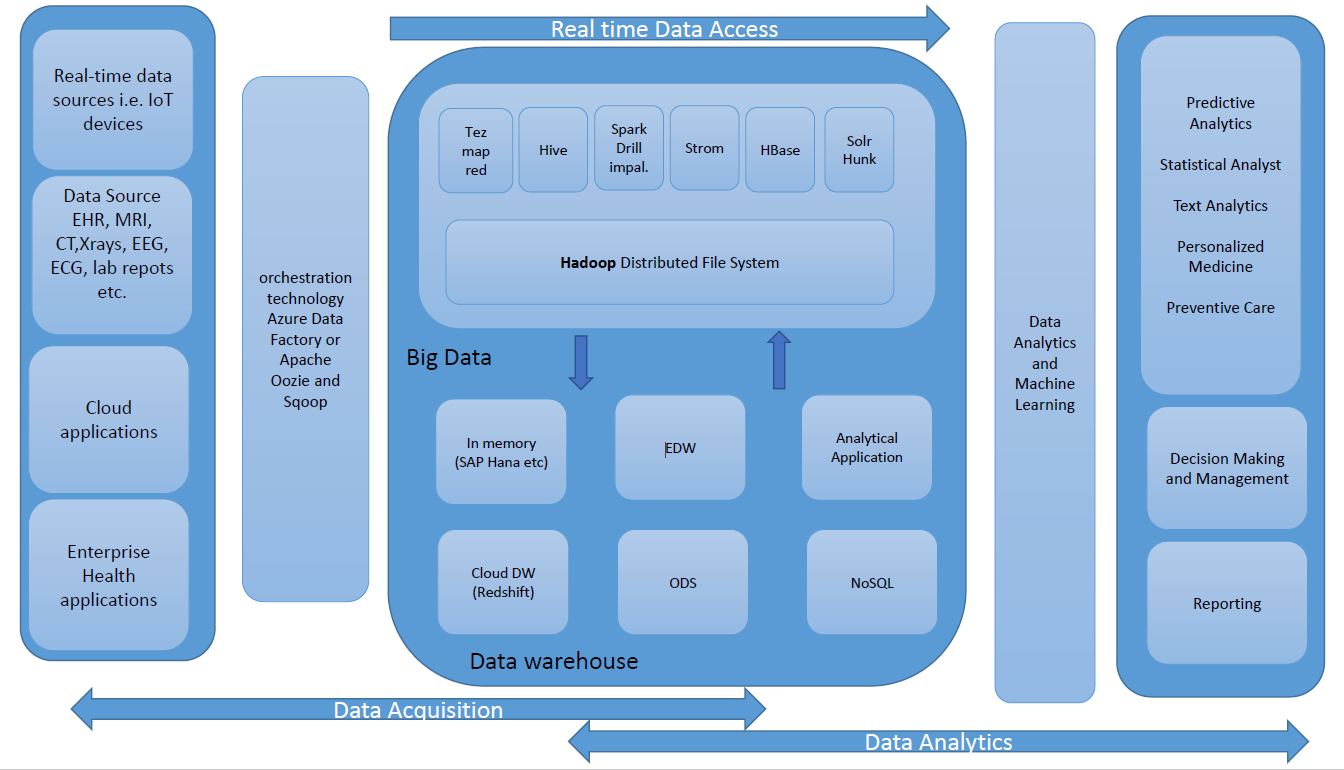}
  \caption{ Architecture of Big Data Analytics Platform}
\label{fig:BigDataArchitectures}       % Give a unique label
\end{center}
\end{figure*}

\begin{figure}[hbtp]
\begin{center}
  \includegraphics[width=8cm,height=4cm]{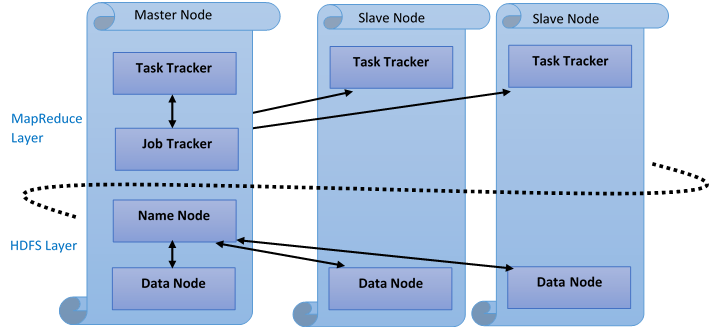}
  \caption{Core Components of Hadoop }
\label{fig:Hadoopcomp}   
\end{center}
\end{figure}

\begin{figure}[hbtp]
\begin{center}
  \includegraphics[width=8cm,height=5.5cm]{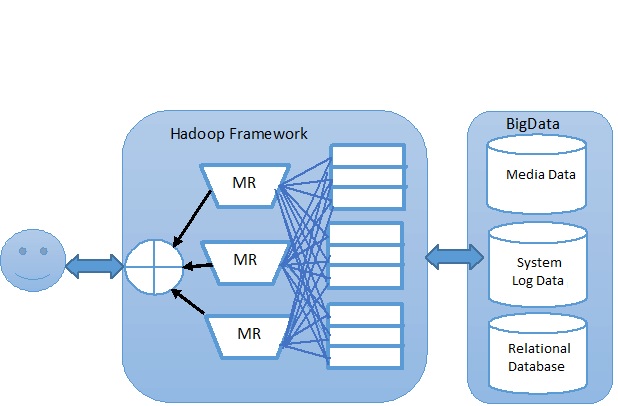}
  \caption{Framework of Hadoop  }
\label{fig:Hadoopframe}  
\end{center}
\end{figure}

Based on the domain abstraction and identification, there are several definitions of big data architectures proposed and developed by researchers for big data analytics. Some the important architectures are Hadhoop, MapReduce~\cite{dean2008mapreduce}, Streaming graph~\cite{stanton2012streaming}, Fault tolerant graph etc.  We present some of the renowned architectures along with its core component comprehensively in detail. One of the major framework on Apache platform is Hadoop developed by Doug Cutting and Apache Lucene. It is a collection of open-source software utilities used for distributed computation, processing and storage of huge data sets or big data. Two architectures or core component of Hadoop are:
\begin{itemize}	
\item Hadoop Distributed File System (HDFS)
\item	MapReduce
\end{itemize}
Succeeding Fig.~\ref{fig:Hadoopcomp} and Fig.~\ref{fig:Hadoopframe}, depicts the core components and basic framework Apache Hadoop.
\subsection{Hadoop Distributed File System (HDFS)}
HDFS \cite{shvachko2010hadoop} is the master-slave architecture intended to run on the commodity hardware. It provide great throughput access to application data. It allows the underlying storage for the Hadoop cluster and enhances healthcare data analytics system by segregating huge expanse of data into smaller one and disseminated it across various servers/nodes. The architecture of HDFS is divided into Name-node and Data-node where Name-node is master and Data node is slave. Documents are stored in the data node having size of 64M that can not be changed. Following Fig.~\ref{hdfsarchitecture} illustrates the architecture of HDFS.
\begin{figure}[hbtp]
\begin{center}
  \includegraphics[width=7cm,height=5.5cm]{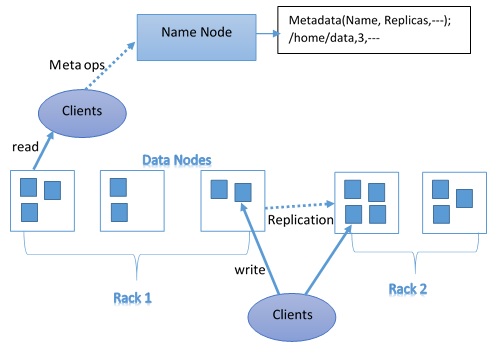}
  \caption{Architecture of HDFS }
\label{hdfsarchitecture}
\end{center}
\end{figure}
According to  Fig.~\ref{hdfsarchitecture}, Client is a HDFS user. Name-node is responsible to manage the name space in the file system. It stores and maintains the files and folders into a file system tree .The Data node is the place where the real data is saved and handles.

\subsection{MapReduce}
Mapreduce is the another cornerstone of Apache Hadoop that is developed in 2004 when Google published a thesis \cite{dean2008mapreduce}. MapReduce is a standard functional programming model that process and analyze . It breaks task into sub-tasks , gathering its outputs and analyze efficiently large datasets in parallel mode. Data analysis and processing employed two steps namely, Map phase and Reduce phase.

\begin{figure}[hbtp]
\begin{center}
  \includegraphics[width=7cm,height=3cm]{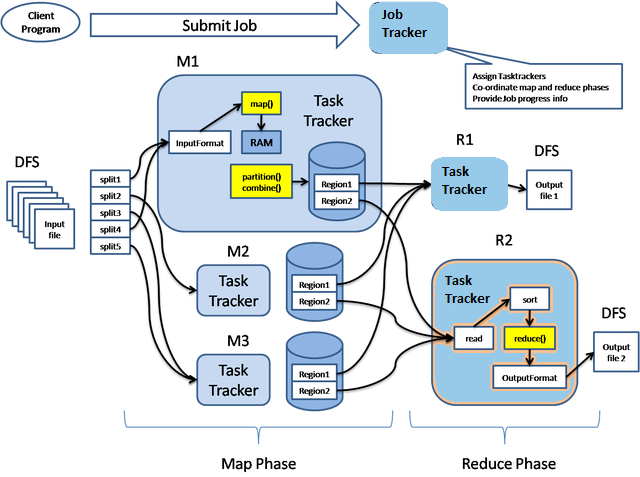}
  \caption{MapReduce Architecture}
\label{mapreducearch}    
\end{center}
\end{figure}

The architecture of MapReduce operation is split into three main components: Client, Job-Tracker and Task-Tracker. Client submit its job to the Job-Tracker in the form of JAR file. Job-Tracker maintains all the jobs that are executed on the MapReduce thus act as master service. Task-Tracker executes the jobs that are assigned by Job-Tracker thus act as slave service. Fig.~\ref{mapreducearch}  demonstrates the generic architecture of MapReduce operation.
\subsection{Apache Hive}

\begin{figure}[hbtp]
\begin{center}
  \includegraphics[width=8cm,height=5cm]{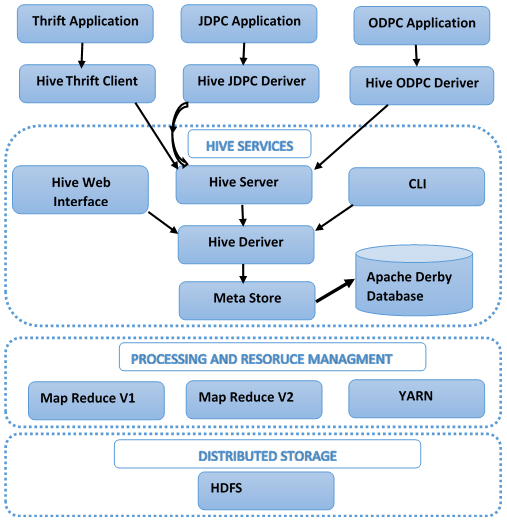}
  \caption{Hive Architecture}
\label{hive}     
\end{center}
\end{figure}

Apache Hive \cite{capriolo2012programming} is a Structured Query Language (SQL) based Extract Transform Load (ETL) and dataware house on Hadoop plateform. It is a run time Hadoop provision framework that works on Hive Query Language (HQL) that converts SQL queries into MapReduce jobs. The main operations performed by Hive are data encapsulation, analyzing, adhoc querying and summarizing large data-sets. 
Apache Hive have four major components: Hive Clients, Services, Processing framework and Distributed Storage. Hive client like Thrift Clients, JDBC Clients, ODBC Clients etc.  can be written in any supportive language like C++, Java, Python etc.Services are used to perform queries. Services of Hive may include command line interface (CLI), Web interface (WI), Hive server, driver, meta-store etc. Queries are processed, executed and managed using internal Hadoop MapReduce framework. Finally the distributed data is deposited in HDFS. The core components are revealed in Fig.~\ref{hive}.
\subsection{Apache HBase}
\begin{figure}[hbtp]
\begin{center}
  \includegraphics[width=8cm,height=4.5cm]{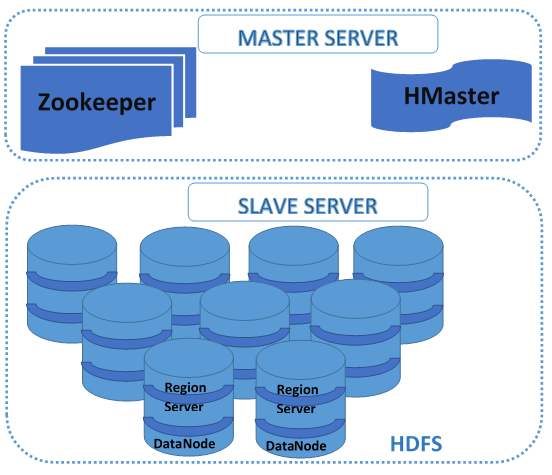}
  \caption{Hbase Architecture}
\label{hbase}     
\end{center}
\end{figure}

Apache HBase  works on non-SQL and non-relational approach. It is a database management approach using column oriented structure lies on the top of HDFS. It used the key/value data that perform read/write operations on large HDFS database. Apache Hbase is categorized into three main components: HMaster Server, HBase Region Server, and Zookeeper. HMaster Server is the main component that manages and monitors  HBase Region Servers, perform database operations using DDL to create, update and delete tables. Hbase tables are divided into several regions that are manage, handle and execute operations through Hbase Region Servers. Hbase is a distributed system that is coordinate by Zookeeper. The components of Apache HBase are depicted in Fig.~\ref{hbase}.
\subsection{Presto}
\begin{figure}[hbtp]
\begin{center}
  \includegraphics[width=8cm,height=5.5cm]{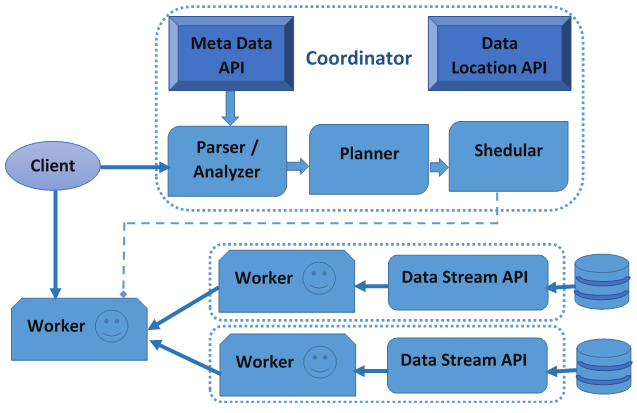}
  \caption{Presto Architecture}
\label{presto}     
\end{center}
\end{figure}

Presto is a distributed structured query language engine that is used to analyzed the large amount of data ranging from in size from gigabytes to petabytes. The architecture of Presto is composed of coordinators and workers. User queries are submitted to the Coordinator that is accountable for planning, executing, scheduling and parsing the queries of Workers. The architecture is explained from the succeeding Fig.~\ref{presto}.
\subsection{Mahout}
Mahout is an apache scheme, objective is to produce unrestricted applications of disseminated and accessible machine learning algorithms that supports healthcare data analytics on Hadoop systems. It is designed to support big data analytics that provide free application on Hadoop platform like applications of distributed and accessible machine learning algorithms. 
 
 \subsection{Avro}
\begin{figure}[hbtp]
\begin{center}
  \includegraphics[width=8cm,height=6cm]{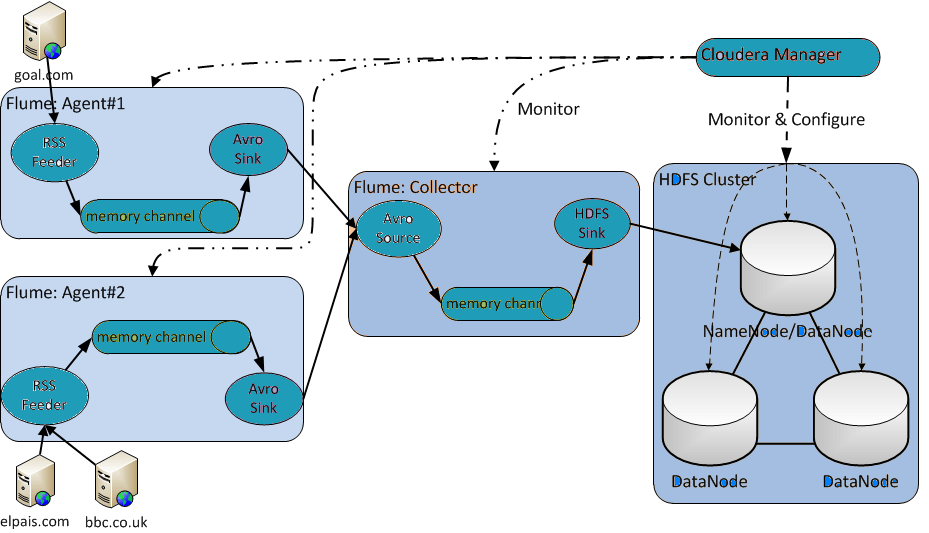}
  \caption{Avro Architecture}
\label{avro}     
\end{center}
\end{figure}

 Avro assists serialization and data encoding that advances structure of data by identifying data types, meaning and scheme. It has the functionalities of serialization and versioning control features. Avro configuration is illustrated from the Fig.~\ref{avro}.
\section{Advantages  of Big Data to Healthcare}
 How big data analytics can improve healthcare? Simple answer to this question is: Analyzing big data can aid healthcare stakeholders to deliver efficient procedures and insights into the patients and their health.  Numerous benefits can be obtained with big data analytics.  Main source of healthcare data are: EHR (Electronic Health Records), LIMS (Laboratory Information Management system), Pharmacy, MDI (Monitoring and diagnostic instruments), Finance (Insurance claim and billing) and hospital resources. With the advancement of data acquisition devices and analytics techniques, data source are getting enriched with newer forms of data i.e. hospitals start to collect Genetic information in EHR as well. Within this vast variety of patient data lies the valuable insights for both patient as well as organizations, which, when applied judiciously can bring in wonderful results.  Potential benefits includes advanced patient care:

\textbf{ Quality of Care:} EHR helps in assembling demographic and medical data such as clinical data, lab test, diagnoses, and medical conditions. By discovering associations and patterns within this data, helps healthcare practitioners to provide quality care, save lives and lower costs.

\textbf{ Disease Prevention:} Spending more on health does not guarantee health system efficiency. The investment in prevention can help to reduce the cost as well as improve health quality and efficiency. Health systems face considerable challenges in endorsing and protecting health at a time when the burden on finances and resources is substantial in many countries. The early detection and prevention of disease plays a very important role in reducing deaths as well as healthcare costs. Thus, the core question are: How can we diminish the level of ill health in the population? And how can we prevent the disease to occur based on early symptoms of patient?

\textbf{Efficiency: } Managing healthcare data using traditional analytical tools is nearly impossible due to the diversity and volume of data. Healthcare stakeholders use big data as a part of their business intelligence strategy to examine historical patient admission rates and to analyze staff efficiency. 

\textbf{Disease Cureness:} Healthcare practices have largely been reactive where the patient has to wait until the onset of disease after which treatment is prescribed which hopefully leads to a cure.However, no two persons in the world would have the same in genetic sequence. Furthermore, environmental factors associated with the onset of the disease are not known., which is the motive why particular medication seems to work for few people but not for others. Since there are millions of things to be considered in a single genome, it is almost impossible to study them comprehensively. On the other hand, big data in healthcare have been revolutionizing the expanse of genomics medicine. Big data analytics can extract hidden patterns, unknown correlations, and insights by exploring large data-sets. Scientists are banking on big data to discover the cure for cancer.

\textbf{Cost:} Healthcare cost can be cut down by analyzing bid data i.e. predictive analytics can helps to detect disease at early stage. Moreover, big data also reliefs in reducing medication errors by advancing economic and administrative performance, and reduce re-admissions. For example, patient groups effected by a disease and are treated with different drug regimens can be compared to determine which treatment plans work best for the same of similar disease which result in saving resources and money.

\textbf{Finding diseases cure:} A particular medication seems to work for a few people but not for others, and there are numerous things to be discovered in a single genome. It isn't feasible to observe all of them in element. however big statistics can help in uncovering unknown correlations, hidden styles, and insights by using analyzing large sets of statistics. through applying machine getting to know, big facts can have a look at human genomes and find the correct remedy or drugs to deal with cancer.

\section{Review Methodology}

The review methodology is the systematic process of finding the relevant literature from different sources. The main objectives of review methodology are:
\begin{itemize}
    \item To deploy the definitions and concepts of Big data in healthcare.
    \item To explore the five sub-disciplines (i.e., medical image processing and imaging informatics, bioinformatics, clinical informatics, public health informatics, medical signal analytics \cite{razzak2019multiclass,razzak2018robust}) that directly or indirectly involve in healthcare and bio-medical.
    
    \item To illustrate the repositories and complex datasets of five sub disciplines.
    \item To determine the big data analytical architectures and techniques in healthcare.
    \item  To discuss the potential advantages and applications of big data in healthcare.
    \item To present the open challenges and research issues of big data in healthcare and the strategies tackling the challenges facing in the domain.
\end{itemize}
	
	he main steps of review methodology are information sources, selection criteria, and search and selection procedure.
%\subsection{Information Sources}
\textit{Information Sources}: The first step in the systematic process of research methodology is to collect the relevant articles. To search the relevant articles we used Google Scholar. We scanned the references to present a thorough review. 
%\subsection{Selection Criteria}
\textit{Selection Criteria}: In second step, we selected the literature on the basis of following inclusion-exclusion criteria:
\begin{itemize}
    \item  Studies were based on articles and reviews
    \item Studies written in English language
    \item Studies related to the big data analytics in healthcare
    \item Studies published from 2000 to 2019
\end{itemize}
	
%\subsection{\textcolor{blue}{Search and selection procedure}}
\textit{Search and selection procedure}: In the third step, we search the studies from the information sources containing the keywords of “big data”, “big data analytics”, “healthcare”, “biomedical” and “healthcare analytics”. As mentioned earlier, our goal is to expand the research in healthcare using five sub-disciplines, we used the additional keywords: “medical”, “medical image processing” “imaging informatics”, “bioinformatics”, “clinical informatics”, “public health informatics”, “medical signal analytics”.  On the basis of initial search criteria, 47,130 papers were found thus we scrutinized the title, keywords and abstract and exclude 28,280 papers. We also perform the screening on the basis of full text reading and exclude 18,020 papers that are irrelevant to the big data or healthcare domain. We ended with 830 papers that are included in this review paper. 

\begin{figure}[htp]
\begin{center}
  \includegraphics[width=8cm]{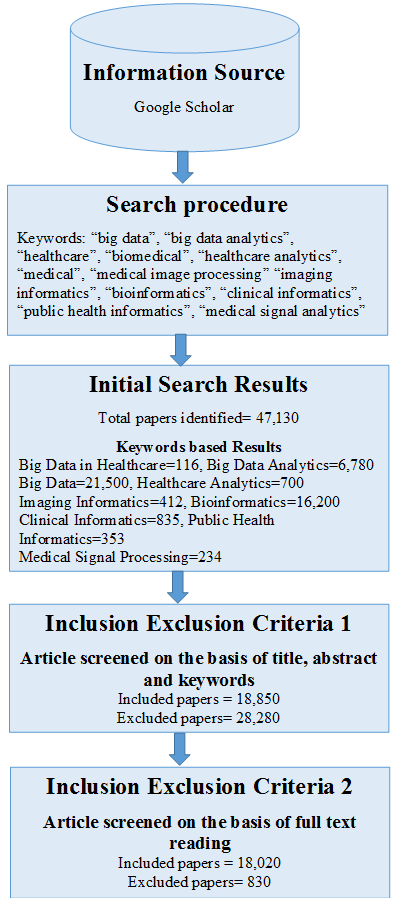}
  \caption{Schematic Process of Review Methodology}
\label{rm}     
\end{center}
\end{figure}

The abstract symbols are used to present    schematic process of review methodology in Fig.~\ref{rm}.

 %disease prevention
\section{Key Application in Healthcare}

 Health professionals, just like business entrepreneurs, are capable of collecting massive amounts of data and look for best strategies to use these numbers to reduce costs of treatment, predict outbreaks of epidemics, avoid preventable diseases and improve the quality of life in general.

Different domains of healthcare and medical care had been proposed in the literature. The general overview, analysis and examples of big data in healthcare analytics was presented in the studies of Raghupathi \cite{raghupathi2014big} and Ward et al. \cite{ward2014applications}. The meaning of big data in healthcare was presented in the literature reviews of Baro et al. \cite{baro2015toward} and Wamba et al. \cite{wamba2015big}.  In 2017, Zhang and Li \cite {zhang2017uses} presented the literature review of specialized healthcare and HIV self-management. Jacofsky \cite{jacofsky2017myths} discussed the pitfalls of analytics related to the physicians from metadata sets in healthcare. Another case study of healthcare analytics was presented in 2018 by Wang et al. \cite{wang2018integrated} that presented IT-enabled procedures, advantages, and capabilities of big data analytics.  Galetsi and Katsaliaki \cite{galetsi2019review} reviewed the articles of big data analytical techniques for healthcare from 2000-2016. 

In this review, we will discuss five sub-disciplines (i.e., medical image processing and imaging informatics, bioinformatics, clinical informatics, public health informatics, medical signal analytics) that directly or indirectly involve in healthcare and bio-medical. As mentioned earlier, we will cover the literature from 2000-2019 that will provide the comprehensive evaluation of big data techniques in healthcare domains. The literature review of five sub-disciplines of healthcare are explained comprehensively in the following subsections.

\subsection{Medical Image Processing and Imaging Informatics}
Medical image processing and imaging informatics are the main applications that play a vital role in healthcare and bio-medical.  One of acceptable use of medical imaging is to detect diseases like tumors detection of brain and lungs, artery stenosis detection, organ delineation detection, aneurysm detection and the diagnosis of spinal deformity and so on. Image processing and machine learning techniques were deployed in these applications for the accurate and effective use of computer-aided medical diagnostics and decision making. In complex healthcare and bio-medical, information is generated, managed, analyzed, exchanged, and represented imaging information using imaging informatics \cite{shirazi2016efficient,razzak2018efficient,naz2017}. 

After the brief introduction, we will elaborate the related work of medical imaging and informatics, techniques and applications deployed in big data healthcare.

\begin{figure}[hbtp]
\begin{center}
  \includegraphics[width=8cm,height=4cm]{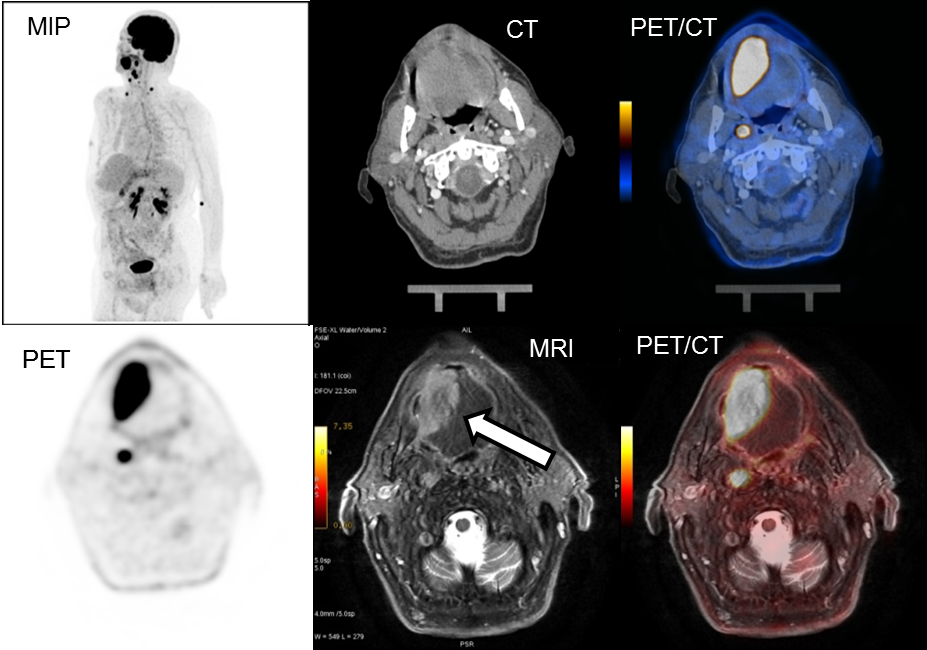}
  \caption{Popular Image Modalities in Healthcare Like CT, MRI, PET images}
\label{fig:images1} 
\end{center}% Give a unique label
\end{figure}

Medical imaging is used in image acquisition. Magnetic Resonance Imaging (MRI), Computed Tomography (CT), photo-acoustic and ultrasound images are used for single dimensional medical data like visualizing the structure of blood vessels \cite{gessner2013acoustic,razzak2018efficient,rehman2019deep}. However for multidimensional medical data like 3d ultrasound, functional MRI (fMRI), Positron-emission tomography (PET) etc. are used as shown in Fig~\ref{fig:images1} \footnote{Available at: http://www.en.nuk.usz.ch/expert-knowledge/PublishingImages/pages/pet-center/PETMR.png}. There are publicly available medical images repositories that contains medical images of patients in different sizes and modalities depicted in the Table~\ref{tab:MedicalImageRepositories}.
\begin{table*}[hbtp]
   \begin{threeparttable}[b]
\begin{center}
\caption{Medical Image Repositories}
\label{tab:MedicalImageRepositories}       % Give a unique label
\begin{tabular}{|p{3cm}|p{1cm}|p{1cm}|p{1.5cm}|p{2.5cm}|p{6.5cm}|}
\hline\noalign{\smallskip}
  Databases & Images & Patients & Data Size & Modalities & Applications  \\
\hline 
Image CLEF Database \tnote{1}  & 306,549 & -- & 316 GB & CT,MRI, PET, Ultrasound & Modality Classification , Visual Image Annotation,  Scientific Multimedia Data Management
\\ \hline
 Digital Mammography database \tnote{2}  & 9,428&2620 & 211 GB & DX &  Computer Algorithm research development for screening
\\ \hline
 Cancer Imaging Archive Database \tnote{3}  & 244,527 & 1010 & 241 GB & CT, DX, CR & image validation of drug response, detection and classification of Lesion,  Diagnostic Image Decision etc.
 \\ \hline
 Public Lung Image Database \tnote{4}  & 28,227 & 119 & 28 GB & CT & Screening Images for identification of lungs cancer
  \\ \hline
 MS Lesion Segmentation  \tnote{5}  & 145 & 41 & 36 GB& MRI & 3D MS Lesion Segmentation Techniques development and comparison
  \\ \hline
  ADNI Database \tnote{6}  &  67,871 & 2851 & 16GB & MRI, PET &    Alzheimer’s disease progression
  \\ \hline

\end{tabular}
  
  \begin{tablenotes}
    \item[1] http://www.imageclef.org/2013/medical
    \item[2] http://marathon.csee.usf.edu/Mammogr aphy/Database.html
    \item[3] https://public.cancerimagingarchive.net/ ncia/dataBasketDisplay.jsf
    \item[4] https://eddie.via.cornell.edu/crpf.html
    \item[5] http://www.ia.unc.edu/MSseg/download .php
    \item[6] http://adni.loni.ucla.edu/data-samples/acscess-data/
     
   \end{tablenotes}
   \end{center}
  \end{threeparttable}
\end{table*}
Shackelford~\cite{shackelford2014system} used fMRI images and single nucleotide polymorphism (SNP) for the classification of schizophrenia and healthy subjects.  They retrieved 87\% classification using hybrid machine learning method.
Chen et al. \cite{chen2010intracranial} introduced a computer-aided decision support system for the treatment of patients with traumatic brain injury (TBI). They predict the intracranial pressure (ICP) level from CT scans images. They combined CT scans images for features extraction, medical records and patient’s demographics. They achieved 70.3\% accuracy, 65.2\% sensitivity and 73.7\% specificity correspondingly.

Yao et al. \cite{yao2014massive} introduced a system for retrieval of medical images based on Hadoop. They applied the local binary pattern algorithm and Brushlet transform for feature extraction of medical images. They implemented MapReduce for storing features in HDFS. They reported highest precision rate of 95.04\% and recall of 92.21\% on brain CT images. They concluded that retrieval efficiency of medical images were improved but retrieval time decreased.

Jai-Andaloussi et al. \cite{jai2013medical} employed the MapReduce for computation and HDFS for storage in content-based image retrieval systems. They used mammography image database and applied Bi-dimensional Empirical Mode Decomposition with Generalized Gaussian Density functions (BEMD-GGD) method and Bi-dimensional Empirical Mode Decomposition with Huang-Hilbert Transform (BEMD-HHT) method. They used Kernal Linear Discriminant (KLD) and euclidean distance. They produced promising results to prove the hypothesis that MapReduce technique can be effectively employed for content-based medical image retrieval.

 Dilsizian and Siegel ~\cite{dilsizian2014artificial} worked on cardiac imaging and medical data by integrating several techniques like data mining, AI, and parallel computing.  Their system use AI and big data for the diagnostic imaging of 55 participating sites from the group of formation of optimal cardiovascular utilization strategies. The system result decreased from 10\% to 5\% in such case.
 
Istephan et al. \cite{istephan2016unstructured} conducted a feasibility study in the epilepsy domain. They used the distributing computation of hadoop clusters. The framework deals with the structured and unstructured medical data.

\subsection{Bioinformatics} 
Bioinformatics is a discipline of sciences which deals with mathematical, computerized and IT-based methods, techniques, algorithms and software tool for capturing, storing, analyzing, compiling, simulating and modeling information of life science and biological data. Role of big data in bioinformatics is to provide efficient data manipulation tools for investigation in order to analyze biological information of patient. Hadoop and MapReduce are currently used extensively used for bioinformatics analytics. 

\begin{figure}[hbtp]
\begin{center}
  \includegraphics[width=8cm,height=5cm]{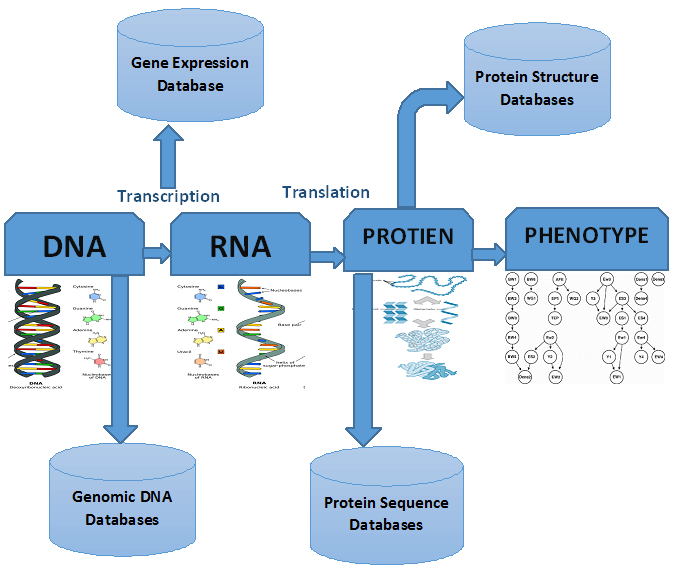}
  \caption{Bioinformatics Types}
\label{fig:images3}       % Give a unique label
\end{center}
\end{figure}

Basically, bioinformatics is the combination of biology and computer science \cite{o2013big}. The biological analysis system analyzes variations at the molecular level. 
The bioinformatics consists of a variety of data types like Genomics (Genes sequencing), RNA, DNA, Proteomics (protein sequencing), gene ontholgoy, protein-protein interaction, pathway data, association network of the disease gene and a network of human disease as shown in Fig~\ref{fig:images3}. 
With the current trends in personalized care, there is an increasing demand to analyze massive size of personalized patient data in a manageable time frame.

The size of bioinformatics' data is increasing exponentially day by day. For example, a single human's sequence of the genome is almost up to 200 GB . A database produced by European Bio-informatics Institution (EBI) has getting double volume after each year \cite{Hirak2015}. \textit{Genomics or Genome sequencing data} is currently being annotated as big data of bioinformatics problem because human genomics consists of 30,000 to 35,000 genes \cite{lander2001initial,drmanac2010human}. Genomics data is usually the data related to gene sequencing, DNA sequencing, genotyping and gene expression etc.\cite{chen2012business,priyanka2014survey} Gene is made of DNA comprising 3 billion pairs of four building blocks or bases known as Adenine, Thymine, Cytosine and Guanine. The single genome has the size of about 3 GB. Genome analysis employing micro-arrays has been profitable in examining traits across a population and widely contributed in treatments of several complicated diseases like bipolar disease, hypertension, rheumatoid arthritis, diabetes, muscular degeneration, coronary heart disease and Crohn’s disease etc. \cite{koboldt2013next}. This genomics information tends to move towards big data analytics.

\begin{table*}[hbtp]
\begin{center}
\caption{Bio-informatics Databases}  
%\cite{chang2010bioinformatics}
\label{tab:Bioinformatics databases}       % Give a unique label
\begin{tabular}{|p{4cm}|p{1.5cm}|p{2.5cm}|p{8cm}|}
\hline\noalign{\smallskip}
 Database & Database Type & Size & Description  \\
\hline
European Molecular Biology Laboratory (EMBL) \cite{kanz2005embl} & DNA Sequences & 185000 organisms & EMBL is the part of an international alliance with DDBJ (Japan) and GenBank (USA). It is used to analyze collection of nucleotide sequences and annotation from sources that are publically available.
\\ \hline
Genetic Sequence Data Bank (GenBank) \cite{bilofsky1988genbank,yao2009mtdna} & DNA Sequence & 15000 DNA and RNA sequences entries & This database contains nucleotide sequences that provide information based on functional and physical contexts of the sequences.

\\ \hline
DDBJ \cite{sugawara2007ddbj} & DNA Sequences & 1880115 entries and 1134086245 bases & This dataset is known as All-round Retrieval for Sequence and Annotation that enable its users to search keywords from Nucleotide Sequence Database Collaboration 
\\ \hline
The GDB Human Genome \cite{gdb1} & Genomics Database & &Public Database of human genes, clones, STSs, polymorphisms and maps
\\ \hline
SWISS-PROTT \cite{boeckmann2005protein,boeckmann2003swiss} & Protein Sequences & 557012 sequence entries, comprising 199714119 amino acids& It contains information of protein variety, function and
associated disorders
\\ \hline 
UniProtKB / TrEMBL & Protein Sequences & & Computer-annotated protein sequence database. It contains sequence translation of coding sequences present in the EMBL/GenBank/DDBJ  
\\ \hline
PROSITE \cite{hulo2006prosite} & Protein Sequences & 1329 patterns and 552 profile entries  & This database contains meaningful biologically signatures that described patterns or profiles
\\ \hline  
PDP \cite{kouranov2006rcsb} & Protein Structure  & 32500 structures & This repository is informative with online reports, summaries, tools and information related to structural genomics initiatives
\\ \hline
BiowareHouse \cite{lee2006biowarehouse} &  Comprehensive Database & & This detailed repository is the integration of the set of databases including ENZYME, KEGG, and BioCyc, and in addition the UniProt, GenBank, NCBI Taxonomy, and CMR databases, and the Gene Ontology 
\\ \hline
\end{tabular}
  \end{center}
\end{table*}

In bioinformatics, protein sequencing and protein-protein interaction are sophisticated problems in functional genomics. This is due to huge number of enormous features in feature vector that is not only cost effective and complex analysis, but also reduces accuracy. Thus feature selection of big data problem is overcome by the method proposed by  Bagyamathi et al \cite{bagyamathi2015novel}. They combined improved harmony search algorithm to improve the accuracy and feature selection. Likewise, another feature selection methodology was introduced by Barbu et al. \cite{barbu2013feature}. They reduced the dimensionality of an instance using annealing technique for big data learning. Similarly, adaptiveness or behavior of big data is predicted by Incremental learning approach. For this purpose, Zeng et al. \cite{zeng2015fuzzy} implemented incremental feature selection method called FRSA-IFS-HIS. They applied fuzzy rough set theory on Hybrid information systems and reported better performance in big data feature selection.

Once the features were extracted and selected, next step is classification or clustering. Classification is the supervised learning procedure of finding a model that describes and discriminates data classes or concepts. The model is used to predict the class label of test instances from already trained instances. Among numerous models described in the literature, linear and non linear density-based classifiers, neural networks, decision trees, support vector machines (SVMs), Naive Bayes, and K-nearest neighbour (KNN) are the most often used methods in numerous applications\cite{mitchell1997machine,duda1973pattern,anzai2012pattern,mohri2012foundations}. In big data analytic, advanced models had been reported in the literature like neural networks approaches, divide-and-conquer SVM \cite{hsieh2014divide}, Multi-hyper-plane Machine (MM) classification model \cite{djuric2013big} etc. for big data parallel and distributed learning. 

Giveki et al. \cite{giveki2012automatic} diagnosed automatic detection of diabetics using weighted SVM on mutual information and modified cuckoosearch. They conducted experiment on diabetics datasets by selecting features from PCA. Haller et al. \cite{haller2012i} classified Parkinson patients by employing SVM. They performed pre-processing using DTI fractional anisotropy data and select most discriminated voxels as features and then classified using SVM. Son et al. \cite{son2010application} predict the heart failure patients by deploying SVM. Likewise, Sumit. Bhatia et. al. \cite{bhatia2008svm} classified heart disease by SVM. They selected optimal feature subset using integer-coded genetic algorithm.

The big data classification and regression is effectively performed using advanced decision tree. In bioinformatics, Jerry. Ye et. al.\cite{ye2009stochastic} implemented Gradient Boosted Decision Trees(GBDT) techniques to distribute and parallelize big data.  Calaway et al. \cite{calaway2016big} estimated efficiency of decision tree on big data by employing rxDTree. Hall et al. \cite{hall1998decision} modified decision tree learning by generating rules for large training data-set.

Clustering is the unsupervised learning that analyzes data objects without labeled responses. To handle big data CLARA \cite{kaufman2009finding}, CLARANS \cite{ng2002clarans} DBSCAN \cite{ester1996density}, DENCLUE \cite{hinneburg1998efficient}, and CURE \cite{guha1998cure}, k-mode and k-prototype methods \cite{huang1998extensions}, PDBCSCAN \cite{xu1999fast}, IGDCA \cite{chen2002incremental}, methods were used in the literature. Literature divulges several bioinfoamtics repositories \cite{kumar2016big} explained in the Table \ref{tab:Bioinformatics databases}.   

Along that there were several techniques and tools employed in bioinformatics for specific task. One of the bioinformatics type is microarray data analysis. Tools used for this type were caCORRECT \cite{stokes2007chip} and omniBiomarker \cite{phan2013omnibiomarker}. For gene-gene network analysis, FastGCN \cite{liang2015fastgcn}, UCLA Gene Expression, Tool (UGET) \cite{day2009disease}, WGCNA \cite{langfelder2008wgcna} tools were used for specific tasks like finding disease associated with genes, parallelism with GPU etc. Several tools had been proposed for Protein -Protein interaction (PPI) that is a complex and time consuming process. NeMo \cite{rivera2010nemo}, MCODE \cite{bader2003automated}, and ClusterONE \cite{nepusz2012detecting}, PathBLAST \cite{kelley2004pathblast} had been developed for PPI analysis. For pathway analysis, GO-Elite \cite{zambon2012go}, PathVisio \cite{van2008presenting}, directPA \cite{yang2013direction}, Pathway Processor \cite{grosu2002pathway}, Pathway-PDT \cite{park2013pathway} and Pathview \cite{luo2013pathview} tools had been employed.

In Protein-Protein Interaction and Protein Sequence, Sequencing data was mapped with the specific genomes for the analysis of various tasks like genotype and expression variation. As DNA sequencing is produced from sequencing machines ranges from millions of data therefore matching with the genomes is one of major task. There are several techniques for the matching of DNA sequence with reference gene.  A parallel computing model for matching genomes is CloudBurst \cite{schatz2009cloudburst}. It use 24 core clusters for evaluation that is 24 times faster in speed than single core system. It has the capability of short read mapping of 7 million reads that improved the scalability of reading huge sequencing data. On the basis of CloudBurst, Contrail \cite{schatz2010contrail} was developed to accumulate hefty genomes and for the identification of single nucleotide polymorphisms (SNP), Crossbow \cite{gurtowski2012genotyping} was prepared. 

A proteomic search engine based on Hadoop distributed framework is Hydra \cite{lewis2012hydra} software package. It is a distributed computing environment that process large peptide and spectra databases to support searching of immense volumes of spectrometry data. It has the fast processing of performing 27 billion peptide scorings on a 43-node Hadoop cluster in approximately             40 minutes. Another query engine for bioinformatics and genomics researchers is SeqWare \cite{d2010seqware} built on Apache HBase \cite{george2011hbase}. Th SeqWare has an interactive interface with genome browsers and tools. It includes loaded U87MG and 1102GBM tumor databases used for the comparison with other prototypes.

There are certain tools used for the error identification of sequencing data. SAMQA \cite{robinson2011samqa} is the error identification tool that provides a scale-able quality for standards for large scale genomic data. ART \cite{huang2011art} can identify three types of errors from sequencing data like base insertion, deletion and substitution. CloudRS \cite{chen2013cloudrs}is a parallel algorithm for error correction. It is based on RS algorithm \cite{gnerre2011high}. For the analysis of data sequencing and genomic analysis, several frameworks and toolkits were developed. CloVR \cite{angiuoli2011clovr,eelmets2011clovr} is a distributed virtual machine package for sequencing analysis that support both local and cloud systems. Another virtual machine tool is CloudBioLinux \cite{krampis2012cloud} that provides 135 bioinformatics packages for analysis. Genome Analysis Toolkit (GATK)\cite{mckenna2010genome,van2013fastq} analyze large sequence and genomics. It based on MapReduce-based programming framework that had been used in 1000 Genomes Projects. BlueSNP \cite{huang2012bluesnp} analyzed 1,000 phenotypes and find association based on R package and Hadoop platform.

\subsection{Clinical Informatics}

\begin{figure}[hbtp]
\begin{center}
  \includegraphics[width=7cm,height=5cm]{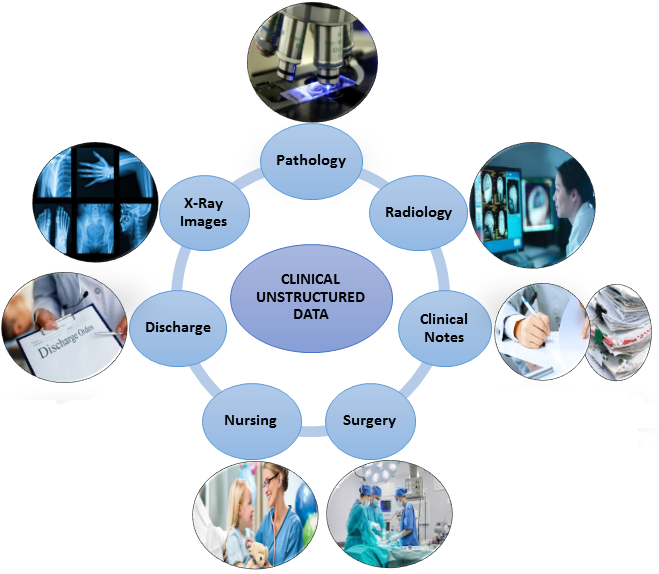}
  \caption{Unstructured Clinical Informatics}
\label{fig:images5}      
\end{center}
\end{figure}

The clinical laboratory is a major source of data related to patients' diseases and health issue. There is approximately 80\% unstructured data like clinical documents, radiology, pathology, patient discharge summaries, diagnostic testing reports, X-ray and radio-logical images and transcribed notes etc.  as shown in Fig~\ref{fig:images5}. Clinical informatics is the study of Information Technology (IT) and healthcare for organizing the patient's clinical data and laboratory test, reports etc. into structured and computerized form to increase data retrieval and extraction efficiently that will assist in evaluations and reports effectively. It divulge the development of electronic health informatics systems for improvement of care and management of patients and sharing of data in seconds using computer and internet. Increasingly laboratory data is being integrated with other data of patient in order to improve the diagnostic process   efficiency, and increase its meaningful use to improve patient outcomes. IT-based systems replace the manual data entry  in records, reports, documents; also save time and cost associated with records, hospital data and reports on daily bases, like billing and schedules of patients \cite{abbott2008globalization}.  However, clinical informatics is currently not practiced in small clinics, hospitals, laboratories in rural and county side areas due to implementation of clinical informatics technology~\cite{bhattacherjee2007physicians}. For boosting the implementation the Electronic Care Records (EHR) system as a clinical informatics in the whole government hospitals in USA, HITEC~\cite{blumenthal2010launching} made some interesting incentives for the medical organizations, hospital and clinics. That the doctors and physicians should use EHR systems for data of patients which they can share with any others and can provide to patients online and or can access anywhere.   

In big data analytics, the first step is to store and manage data in some structured form.  Clinical data is store to observe the information of patients, hospitals and other relevant structured and unstructured record. It can be than used to settle on clinical decision, assessing patients and make treatment plans. Data warehouses and relational databases are the traditional and structured methods to store and retrieve data. However, to use clinical data, it is first transformed and classified when it is integrated from multiple sources \cite{bakshi2012considerations,herodotou2011starfish}.
A detailed systematic review paper  is published in \cite{Buntin2011} till 2011. We here presents the further related work.  Dutta et al. \cite{dutta2011distributed} stored EEG data using Hadoop and HBase in data warehouses.   Jin et al. \cite{jin2011distributed} analyzed and stored distributed EHR data using big data tools like Hadoop HDFS and HBase. Similarly, Nguyen et al. \cite{nguyen2011hbase} stored signal clinical data using HBase.  Jayapandian et al. \cite{jayapandian2013cloudwave} and Sahoo et al.\cite{sahoo2013heart} developed a system named 'Cloudwave' for storing and querying EEG clinical data that is voluminous. Mazurek \cite{mazurek2014applying} stored unstructured data in Not Only Structured Query Language (NoSQL) repositories to provide fast processing speed and data mining capabilities. For this purpose, relational and multidimensional technologies were combined with NoSQL. 

Clinical data is often retrieved and shared interactively for data integration and knowledge sharing, so the cloud computing was the usually consider for this purpose. Bahga and Madisetti \cite{bahga2013cloud} proposed a system based on cloud approach for inter-operable EHRs. Chen et al.\cite{chen2013translational} translated the informatics aspects of present and future using cloud computing. For multi-site clinical traits, the interactions of researchers were enhanced by the conceptual software architecture developed by Sharp \cite{sharp2011application} using cloud approach. Clinical data is analyzed to predict the disease, risk, diagnosis, and progression. Literature divulges a lot of data analysis strategies for the prediction of clinical record. One of the predictive modeling platform was "PARAMO" designed by Ng et al.\cite{ng2014paramo}   for analyzing EHR and the generation and reuse of clinical data. using a Hadoop cluster. They analyzed the EHR  from 5,000 patients to 300,000 patients and reported promising time effective results. Chawla and Davis \cite{chawla2013bringing} formulated the framework for patient-centered to explained the big data approaches for personalized medicine. Similarly, the big data for perioperative medicine were illustrated by Abbott \cite{abbott2013big}. Zolfaghar et al. \cite{zolfaghar2013big} implemented big data techniques for the predictive model. They conducted an experiment on patient data of "National Inpatient Dataset and the MultiCare Health System" for the congestive heart failure. They reported the maximum accuracy upto 77\% and recall upto 61\%, respectively.  Rangarajan et al. \cite{rangarajan2015scalable} proposed data lake architecture that used HDFS for data storage. Similar health conditions of patients were clustered using K-means. From each cluster, the successful recommendation was found by deploying SVM. Wang and Hajli \cite{wang2017exploring} examined 109 case description of 63 healthcare organizations. They modeled the big data analytics for business transformation using RBT theory and capability building view in the model. Each case occurrences along with pair-wise connections, constructs and path-to-value chains were used to find business value.

\begin{table*}[hbtp]
  \begin{threeparttable}[b]
\begin{center}
\caption{Clinical informatics Databases} 
%\cite{chang2010bioinformatics}
\label{tab:Clinicaldb}       % Give a unique label
\begin{tabular}{|p{4cm}|p{1.7cm}|p{11cm}|}
\hline\noalign{\smallskip}
 Database & Database Type & Description   \\
\hline Texas Inpatient Public Use Data File (PUDF) \tnote{1} & Structured EHR  & This dataset contains record of patients, hospitals,admission
type/source, claims, admit day and discharge details. In 2017 dataset contains 699 hospitals,  776,554 base date records, 12,486,488 charges date records in First quarter. In Second quarter there were 694 hospitals, 761,921 base date records and 11,985,920 charges date records.
\\ \hline
Multi-parameter Intelligent Monitoring in Intensive Care II (MIMIC-II) Clinical Database \cite{saeed2011multiparameter} & Structured EHR & This dataset encompasses detailed clinical data, including physiological wave forms and records subsets from minute-by-minute. It contains 32,536 subjects with 40,426 ICU admissions and 25,328 intensive care unit stays. 
\\ \hline
Patient Discharge Data By Admission Type \tnote{2} & Unstructured  &  Dataset contains the information of inpatient discharges by type of admission for each California hospital for years 2009-2015 containing 9,322 entries.
\\ \hline
Framingham Heart Study Database \tnote{3} & Structured EHR & It is a genetic dataset for cardiovascular diseases like Heart. It include 5,209 men and women having age between 30 and 62 years. 1948, participants had been assessed every 2 years 
\\ \hline
Basic Stand Alone (BSA) Medicare Claims Public Use Files (PUFs) \tnote{4} & Unstructured & CSV format that contain non-identifiable claim-specific information and are within the public domain.
\\ \hline 
Nationwide Inpatient
Sample \tnote{5} &  Structured EHR &  This dataset contains discharge information including diagnosis, procedures, status,
demographics, cost and length of stay. It comparises 1051 hospitals of 45
states.    
\\ \hline
i2b2 Informatics for
Integrating Biology \& the
Bedside \tnote{6} & Unstructured Clinical Data &  Clinical notes used for clinical NLP challenges like deidentification, Smoking, Obesity, Medication, Relations and co-reference challenges
\\ \hline 

\end{tabular}
 
  \begin{tablenotes}
    \item[1] http://www.dshs.texas.gov/thcic/hospitals/Inpatientpudf.shtm
    \item[2] https://data.chhs.ca.gov/dataset/patient-discharge-data-by-admission-type/resource/460bd2e8-3b0e-4a41-b2a6-1044f7c82178
    \item[3] https://epi.grants.cancer.gov/pharm/pharmacoepi.html
    \item[4] https://www.cms.gov/Research-Statistics-Data-and-Systems/Downloadable-Public-Use-Files/BSAPUFS/index.html
    \item[5] https://www.hcup-us.ahrq.gov/news/exhibit\_booth/nis\_brochure.jsp
    \item[6] https://www.i2b2.org/
    
   \end{tablenotes}
    \end{center}
  \end{threeparttable}
\end{table*}

\subsection{Public Health Informatics}

Informatics is an "Applied Information Science". It synthesizes the practices and theories of information technology, computer science, management sciences and behavioral sciences into concepts, tools and methods for implementing information systems into health for public. Informatics uses to transform raw data into information effectively according to requirement of users. healthcare informatics researches is a scientiﬁc attempt that improve both health service organizations’ performance and patient care outcomes as shown in the following Fig.~\ref{fig:imagespub}.
\begin{figure}[hbtp]
\begin{center}
  \includegraphics[width=7cm,height=4cm]{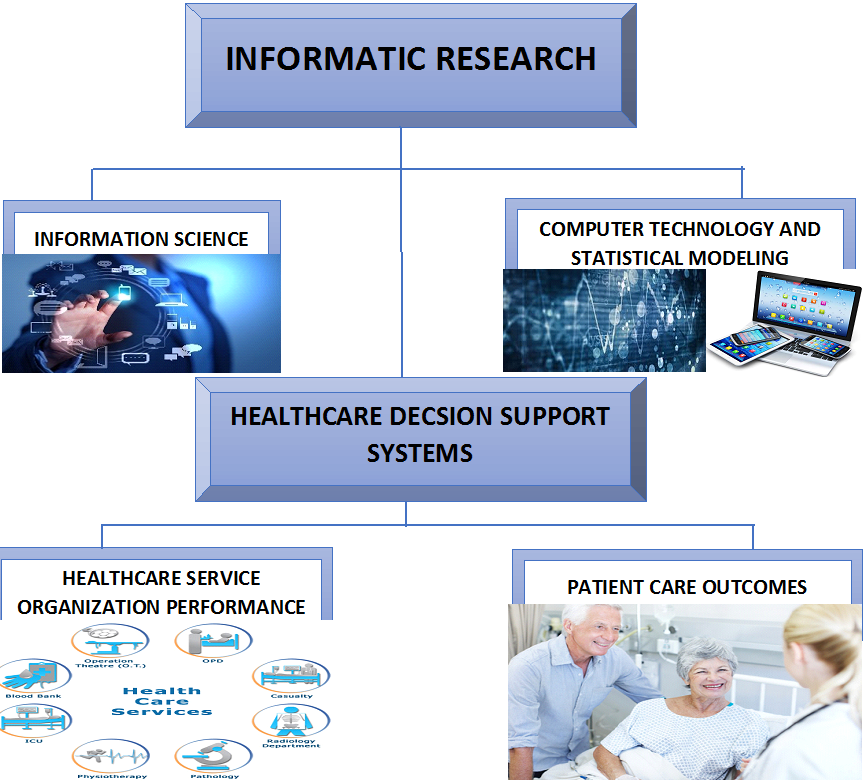}
  \caption{Healthcare Informatics Researches}
\label{fig:imagespub}       % Give a unique label
\end{center}
\end{figure}
Public healthcare is determined through Epidemiology. Epidemiology is the study of analyzing how frequently diseases arise in different groups of people and why. Epidemiological information is used to formulate and evaluate techniques to prevent illness. This information is also serve as a guideline to the management of patients in whom disease has already evolved. Traditionally, epidemiology has been based on data collected by public health agencies through health personnel in hospitals, doctors' offices, and out in the field. 

The healthcare mechanism is the usual first line of reaction to clinical activities, whether of large or less severity. Informatics are used to figuring out sentinel occasions and leading to analysis can keep away from doubtlessly devastating effects. An example of response is \textit{war on cancer} announced in 1973 when the programmers of National Institutes of Health feed the data from registries to the information system entitled with Surveillance Epidemiology and End Results (SEER) system. This system provide the information to the public health planner and epidemiologists to analyze the distribution of cancer throughout the population~\cite{hankey1999surveillance}. After 3 many years of monitoring and evaluation, Age-adjusted mortality rates as a consequence of cancer were dropping step by step since the early 1990, with important development in areas including lung most cancers reflecting fulfillment in public health efforts aimed at controlling precipitants to the disease~\cite{hiatt1999new}. 

Another example of that capacity can be seen inside the response to the 2001 bio-terrorism assaults. During September 2001,anthrax spores had been traced to postal facilities in Trenton, New Jersey and Brentwood, Washington. Epidemiologists face dadaunting venture: the new Jersey facility was a facility of 281,387 square ft, staffed by 250 employees according to shift and processing over 2 million items of mail in line with day~\cite{zubieta}. Informatics helped to become aware of the those who could have been exposed to anthrax,monitored the screening system,and recorded who obtained antibiotics and distribution of recognized cases and known deaths. Further analytical strategies and significant healthcare researches were explained in \cite{wan2006healthcare,revere2007}.

%In recent years, however, novel data sources have emerged where data are frequently collected directly from individuals through the digital traces they leave as a consequence of modern communication [2] and an increased use of electronic devices.

In latest years, innovative data sources have introduced that are used to collect data in a second from individuals directly using electronic devices. Social media change the life of society and make global World. The exponential amount of data is produced daily. Big data is produced from Public Health (PH) information and can be generally characterized as big data.  Public healthcare data is collected, analyzed, assured and accessed so that big data analytics techniques are deployed to extract hidden informative patterns. Public or social media information is further used to predict, monitor and diagnosis of diseases i.e. efficient and effective use of PH data determines the extent to which societal health concerns can be determined. Literature divulges several survey papers based on data mining \cite{herland2014review,kamesh2015review}, deep learning \cite{ravi2017deep,miotto2017deep} and other \cite{aziz2017review}. We here presents some of the public healthcare work using social media. The data-sets for public healthcare data corpus are explained in the Table \ref{tab:Publicdb}. 
\begin{table*}[hbtp]
\begin{center}
\caption{Public Health Databases}  
%\cite{chang2010bioinformatics}
\label{tab:Publicdb}       % Give a unique label
\begin{tabular}{|p{3cm}|p{2.5cm}|p{3cm}|p{8cm}|}
\hline
%\noalign{\smallskip}
 Database & Database Type & Size & Description  \\

\\ \hline
Ohio Hospital Inpatient/Outpatient Database & Public Patients Records  & 35 million patient records per year & This repository contains hospital record such as number of admissions, discharges, stay length, transfers, number of patients with specific codes
\\ \hline
Behavioral Risk Factor Surveillance System (BRFSS) \cite{centers2015behavioral} & Survey System & 50 states data of more than 400,000 adult interviews each year & This system contains record of mental illness, smoking, alcohol, lifestyle (diet, exercise) and diseases (diabetes, cancer) etc.

\\ \hline
Surveillance Epidemiology and End Results (SEER) Program \cite{hayat2007cancer} & Cancer Dataset & 7.7M cases and more than 350,000 cases are added each year & This program contains survival data from population-based cancer registries covering approximately 28\% US population.
\\ \hline
PatientsLikeMe  Online Patient Network Database  \cite{smith2008patientslikeme} & Online Patient Network Database & more than 200,000 patients and is tracking 1,500 diseases& This data corups contains information of disease-specific functional scores, sympotms etc. through which people having same symptoms connect with each others.
\\ \hline
Human Mortality Database \cite{wilmoth2010human}& Public Mortality &39 countries or areas& This database contains information about population and mortality in detail along with Birth, death, population size by country.
\\ \hline
\end{tabular}
  \end{center}
\end{table*}

Young et al.\cite{young2014methods} gathered 553,186,016 tweets from the Twitter. They extracted more than 9,800 keywords and geographic annotations that contains HIV risk words. They revealed that social media monitor global HIV occurrence and concluded that positive correlation of greater than 0.01 was retrieved between HIV-related tweets and HIV cases. 

Hay et al. \cite{hay2013big} facilitated public health surveillance using online social media combined with epidemiological information. They developed atlas for real-time disease monitoring.  

Nambisan et al.\cite{nambisan2015social} detected depression from messages and tweets of social media thus big data analytic tools were used to extract the hidden valuable patterns for detecting mental disorders. They concluded that behavioral and emotional patterns in messages showed the symptoms of depression.

Tsugawa et al.  \cite{tsugawa2013estimating} implemented multiple regression models to detect the depressive tendencies. They extracted frequency of words form messages and Twitter from the popular micro-blogging services to detect depression and achieved a correlation of approximately 0.5. Park et al. \cite{park2012depressive} analyzed depression of 60 participants from their activities on tweeter from sentiment words of depressed users. Another contribution by the same author was to detect the symptoms of depressive users through Facebook \cite{park2013activities}. Choudhury et al. \cite{de2013predicting,de2013social} developed a large dataset from Twitter posts using crowd sourcing methodology. They implemented the probabilistic model to indicate the depression level form social media. In \cite{de2013p} quantified postpartum changes and depression of 376 mothers from Twitter posts. Similarly, same authors in \cite{de2014characterizing} detected and predicted the onset of post-partum depression of 165 mothers through Facebook shared data. Sadelik et al. \cite{sadilek2012modeling} predicted infectious diseases through the social network. They used 1000 Twitter messages related to healthcare. They applied statistical models on geo-tagged postings made on Twitter for prediction of diseases that cause an infection like flu etc. Digital media is widely used to improve healthcare monitoring and its effectiveness. Ginsberg et al. \cite{ginsberg2009detecting} used Trends models and  search queries on Google to detect influenza and flue like diseases.  One of the most earlier comprehensive review paper of public healthcare informatics using social media was presented by Hagg et al. \cite{hagg2018emerging}

\subsection{Medical Signal Analytics}
Nowadays technology is advancing rapidly that provide effectiveness in every walk of life, especially in healthcare. Currently, healthcare systems use a variety of continuous monitoring devices that generate signals. Physiological signal monitoring devices and Telemetry devices are pervasive \cite{belle2015big} because these devices improve healthcare management and patient healthcare \cite{bodo2013multimodal,hu2014identification}. These devices use discretized or physiological waveform data and generate alert mechanisms in case of an overt event. There are certain issues in medical signals that tend to move towards big data. The most notable obstacle is volume and velocity of continuous and high-resolution multitude monitors connected to each patient. The generated alarm systems are unreliable and cause alarm exhaustion for both caregivers and patients \cite{ref10,graham2010monitor}. The primary failure of these systems are due to the relay on single sources of information.

\begin{figure}[hbtp]
\begin{center}
  \includegraphics[width=7cm,height=6cm]{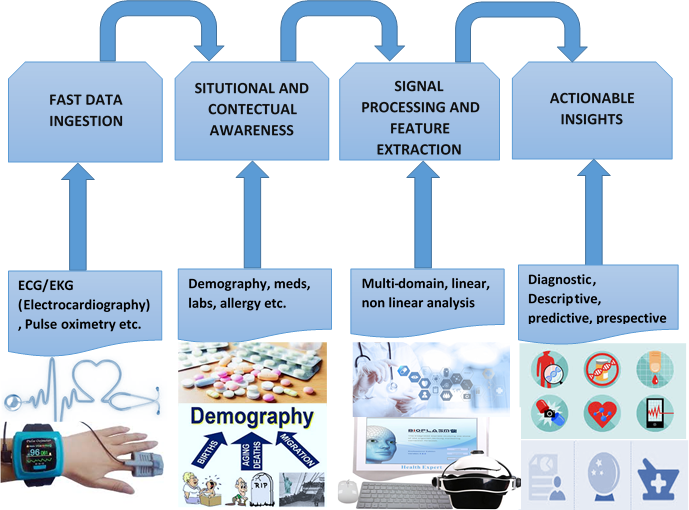}
%  caption is below the figure
\caption{Generalized Work-Flow of Streaming Healthcare}
\label{fig:workflow}       % Give a unique label
\end{center}
\end{figure}

The first step in streaming data analytics in healthcare is to the acquisition of signals. It is usually rare to store the streaming signals from continuous acquisition devices. However to access the live streaming data from devices is one of the foremost tasks for big data analytics applications. As there are many challenges poses to healthcare systems during streaming data collection like network bandwidth, scalability, and cost \cite{mccullough2010effect}. Thus Research communities are developing continuous monitoring technologies \cite{ahmad2009continuous} to capture live monitor signals. Next step is to store the signals data from monitoring devices using Big Data analytics tools like HDFS, MapReduce, and MongoDB \cite{adrian2013mongodb,kaur2015managing} etc. Medical data including signals is complex due to interconnected and interdependent data among several sources. Thus, data is integrated and aggregation techniques are deployed for effective performance \cite{santos2011enabling,berndt2001healthcare}. The workflow of generalized streaming healthcare is depicted in Fig.\ref{fig:workflow}. The most notable data repositories containing signals information in healthcare is explained in Table \ref{tab:Signaldb}.

\begin{table*}[hbtp]
   \begin{threeparttable}[b]
\begin{center}
\caption{Medical Signal Databases}  
%\cite{chang2010bioinformatics}
\label{tab:Signaldb}       % Give a unique label
\begin{tabular}{|p{4cm}|p{2cm}|p{3cm}|p{7cm}|}
\hline
%\noalign{\smallskip}
 Database & Database Type & Size & Applications  \\
\hline
The  Multi-parameter Intelligent Monitoring in Intensive Care II (MIMIC-II) Clinical Database \cite{saeed2011multiparameter} & Structured EHR & 32,536 subjects with 40,426 ICU admissions and 25,328 ICU stays & It contains comprehensive clinical data, including physiological waveforms and minute-by-minute records subsets.
\\ \hline
MIMIC-III Database \cite{johnson2016mimic} & hospital Database &38,597  patients,  49,785 hospital admissions  & This data corpus is inforamative with vital signs, laboratory measurements,medications, imaging reports, details of observations , fluid balance, diagnostic codes, procedure codes, stay length of hospital, survival data, etc.
\\ \hline
The ECG-ID Database \tnote{1}  & Signals Database & 90 persons, 310 ECG recordings & EEG signal recordings each have 10 annotated beats, digitized at 500 Hz with 12-bit resolution and recorded for 20 seconds.
\\ \hline
CinC Challenge 2000 data sets \cite{goldberger2000physiobank}& EEG signal based database & 583 megabytes,  70 records & This dataset contain EEG signals of 70 records, used 35 records for learning set and 35 for testing
\\ \hline
MIT-BIH Polysomnographic Database \cite{goldberger2000physiobank} &  Physiologic Signals Database  & 18 records, each have 4 files & This database is the collection of recordings of multiple physiologic signals during sleep
\\ \hline 
EEG Motor Movement/Imagery Dataset \cite{goldberger2000physiobank} & EEG Signals Database&109 volunteers,1500 recordings  &  Two minutes EEG recordings, 64-channel EEG were recorded using the BCI2000 system 
\\ \hline
American Heart Association (AHA) \tnote{2} & EEG Signals Database & 80 recordings &ECG recordings of 80 two-channel records digitized at 250 Hz per channel with 12-bit resolution with range of 10 mV.
\\ \hline 
\end{tabular}

\begin{tablenotes}
    \item[1] https://www.physionet.org/pn3/ecgiddb/
    \item[2] https://www.physionet.org/physiobank/database/ahadb/
   \end{tablenotes}
   \end{center}
  \end{threeparttable}
\end{table*}

After introducing medical signal analytics, we will present some of the related work of Big Data analytics in medical signaling.  Han et al. \cite{han2006infrastructure} developed a patient care management system using a scalable infrastructure. This system combined static and continuous data from monitored ICU devices. It analyzed and mined medical data in real time. 

Bressan et al. \cite{bressan2012trends} implemented an architecture for neonatal ICU. It used data of EEG monitors, infusion pumps and cerebral oxygenation monitors. Their proposed system provide effective decision system for clinics.

Lee and Mark \cite{lee2010hypotensive}conducted experiment on MIMIC II database for therapeutic intervention to hypotensive episodes. Their system predict intensive care based using blood pressure and cardiac time series data.

Sun et al.\cite{sun2010system} also used MIMIC II database to extract the physiological waveform data along with clinical data. They selected cohorts and find the similarity of patients from them that is beneficial for healthcare. The similarity is used for the treatment of similar diseases and deducted effective decisions from them. Another study on MIMIC II database is to detect the cardiovascular instability in patients at an early stage. For this purpose Cao et al. \cite{cao2008predicting} developed a system that combined multiple waveform data from MIMIC II corpus.

Roux et al.\cite{le2014consensus} discussed the neuro-critical care of the patient's disorders using different physiological monitoring systems. They provide a platform for the researchers with guidelines by examining the potentials and implications of neuro-monitoring.

Rajan et al. \cite{rajan2018internet} used a multi-channel signal acquisition method for the development of physiological signal monitoring system using NI myRIO connected with the wireless network. They also used the Internet of Things( IOT) techniques for better performance in healthcare.

Zhang et al. \cite{zhang2018sensor} recognized the Lung cancer using sensor-based wrist pulse signal processing with the technique of cubic support vector machine (CSVM). They implemented iterative slide window (ISW) algorithm for signal segmentation and extract 26 features. Using these strategies, they achieved 78.13\% accuracy. Nanda et al. \cite{nanda2015quantitative} distinguished between essential tremor and Parkinson’s tremor using non-invasive recording techniques. They employed Neural Network for the classification of tremor sEMG signals and achieved 91.66\% accuracy.

\section{Key Findings}
This survey presents the emerging landscape of big data and analytical techniques in the five sub-disciplines of healthcare. We present various domains of healthcare in which big data technology has played a significant role in modern-day healthcare revolution, as it has totally changed the perception of people about healthcare activities. Big data analytical techniques deployed in five sub-disciplines such as, medical image processing and imaging informatics, bioinformatics, clinical informatics, public health informatics, medical signal analytics are explained comprehensively that draws an integrated depiction of how distinct healthcare activities are accomplished in a pipeline to facilitate individual patients from multiple perspectives. The existing reviews did not provide the detailed explanation in multiple sub-disciplines of healthcare. There is no comprehensive evaluation of studies in the existing reviews. 

The existing studies discussed the different sources of healthcare for big data such as pharmaceutical firms, healthcare providers, diagnostic companies, laboratories, not-for-profit organizations insurance companies and web-health portals \cite{rouse2014,mohammed2014applications,swan2013quantified,ward2014applications,huang2016path,bradley2013implications}. The big data techniques used for the analysis of healthcare data are machine learning, data mining, cluster analysis, pattern recognition, neural networks, deep leaning and spatial analysis. Most of the studies processed the patient data using Hadoop and its tools, but they are batch processing tools \cite{wang2011drug,hung2013implementation,wang2012parallel,meng2011ultrafast,peek2014technical}. There are some studies that used newer tools like Spark, Storm and GraphLab etc. for the processing of real time and streaming data \cite{peek2014technical}. Most of the studies discussed the applications of big data analytics in different fields of healthcare like personalized medicine, clinical decision support, clinical operations optimization and cost effectiveness of healthcare. It can be showed that healthcare analytics improve the quality and early identification of patients. There are researches related to diabetes, gynecology, oncology, cardiovascular diseases and so on that enable save time and cost \cite{wang2018integrated,maia2017big,wong2015big,viceconti2015big,geerts2016big,el2016perspectives}.

\begin{figure*}[hbtp]
\begin{center}
  \includegraphics[width=15cm,height=8cm]{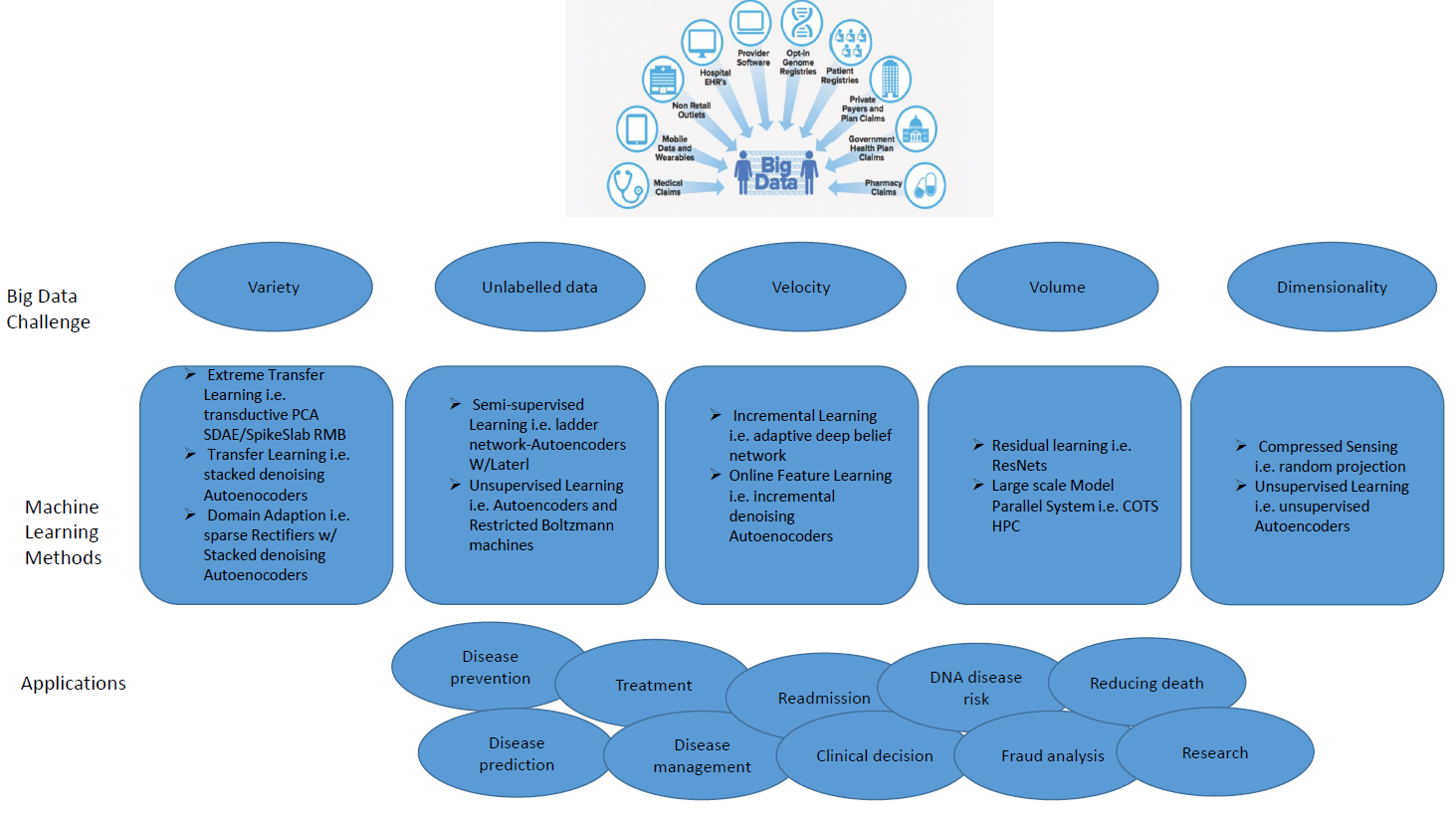}
  \caption{ Deep Learning Architecture for Big Data Analytics}
\label{fig:BigDataArchitectures}       % Give a unique label
\end{center}
\end{figure*}

With the rapid increase of publications in biomedical and healthcare industry, we have conducted the detailed review regarding healthcare analytics in five sub-disciplines. We summarized the usability studies of each discipline in Table \ref{tab:lit}, including image visualization, image classification, image retrieval, data and workflow sharing, data analysis, feature selection, bioinforamtics classification and clustering, micro-array data analysis, protein-protein interaction, pathway analysis, protein sequencing, query and search engine, error identification of sequencing data, storage and retrieval of EHR, treatment recommendation, business transformation, disease prediction, diagnosis and progression, data security, infectious disease surveillance, population health management, mental health management, chronic disease management, signal acquisition, signal storing from monitoring devices, signal integration and aggregation respectively.  
It is concluded from this survey, that bioinformatics is one of primary discipline in which big data analytics is currently evolved and playing a scientific role, due to the complex and massive bioinformatics data. There are a lot of tools, techniques and platforms for bioinformatics used to analyze biological, genomics, proteins and gene sequencing data. However, there is less potential of big data applications in other disciplines such as, in medical imaging informatics, clinical informatics, public health informatics and medical signal analytics.

\begin{table*}[hbtp]
\begin{center}
\caption{Comparative analysis of the literature}
\label{tab:lit}       % Give a unique label
\begin{tabular}{|p{5cm}|p{5cm}|p{6cm}|}
\hline
  Healthcare Discipline & 	Big Data Analytical Technique & 	Studies  \\
\hline

\multirow{4}{*}{\begin{tabular}{@{}c@{}c@{}c@{}@{}c@{}}   Medical Image  Processing  and \\ Imaging Informatics \\   \end{tabular}} & Image Visualization & \cite{gessner2013acoustic,lu2019image,karmonik2018workflow,glemser2018new,yu2018joint,jorge2019challenges} \\ 

 & Image Classification & \cite{shackelford2014system,khan2019novel,lakshmanaprabu2019optimal,rehman2019deep}  \\
 
 &  Image Retrieval &  \cite{yao2014massive,jai2013medical,grace2014medical,yang2015accessing,markonis2012using} \\
 
 & Data and Workflow Sharing & \cite{benjamin2010shared,costa2012telecardiology,ross2011images,ross2011images} \\
 
 & Data Analysis & \cite{dilsizian2014artificial,markonis2012using,wang2011hadoopgis} \\
\hline
\multirow{9}{*}{\begin{tabular}{@{}c@{}c@{}c@{}@{}c@{}}   Bioinformatics \\   \end{tabular}} & Feature Selection & \cite{bagyamathi2015novel,barbu2013feature,zeng2015fuzzy,zou2016novel,tadist2019feature,lualdi2019statistical} \\ 

& Classification  & \cite{giveki2012automatic,bhatia2008svm,ye2009stochastic,calaway2016big,haller2012i,hall1998decision,zou2016novel,david2019classification,devi2019big}
\\ 

& Clustering & \cite{kaufman2009finding,ng2002clarans,ester1996density,hinneburg1998efficient,xu1999fast,chen2002incremental,patel2019big} \\ 

& Microarray Data Analysis & \cite{stokes2007chip,phan2013omnibiomarker,liang2015fastgcn,day2009disease,langfelder2008wgcna} \\

& Protein-Protein Interaction & \cite{rivera2010nemo,bader2003automated,nepusz2012detecting,kelley2004pathblast}
\\
& Pathway Analysis & \cite{zambon2012go,van2008presenting,yang2013direction,grosu2002pathway,park2013pathway,luo2013pathview}
\\
& Protein Sequencing & \cite{schatz2009cloudburst,schatz2010contrail,gurtowski2012genotyping}
\\
& Protein Query and Search Engine & \cite{lewis2012hydra,d2010seqware,george2011hbase}
\\
& Error Identification of Sequencing Data & \cite{robinson2011samqa,huang2011art,chen2013cloudrs,gnerre2011high,angiuoli2011clovr,eelmets2011clovr,krampis2012cloud,mckenna2010genome,van2013fastq,huang2012bluesnp}
\\
\hline

\multirow{5}{*}{\begin{tabular}{@{}c@{}c@{}c@{}@{}c@{}}   Clinical Informatics \\   \end{tabular}}  & Storage of EHR & \cite{blumenthal2010launching,bakshi2012considerations,herodotou2011starfish,dutta2011distributed,jin2011distributed,nguyen2011hbase,jayapandian2013cloudwave,sahoo2013heart,mazurek2014applying,rangarajan2015scalable} \\
& Retrieval of EHR & \cite{goli2016effective,sharp2011application} \\
& Interactive data retrieval for Data Sharing & \cite{deb2001multi,bahga2013cloud,sultana2014cloud,bahga2013cloud,chen2013translational,he2012toward,wang2014large} \\
 & Treatment Recommendation & \cite{chen2010intracranial,chen2018disease,dilsizian2014artificial,istephan2016unstructured} \\
 & Business Transformation & \cite{wang2017exploring,wang2018big,wang2018integrated,gupta2016toward,wang2017business,kim2017identifying} \\
  & Disease Predication, Diagnosis and Progression & \cite{ng2014paramo,chawla2013bringing,abbott2013big,zolfaghar2013big} \\
  & Data Security & \cite{schultz2013turning,sobhy2012medcloud,lin2015cloud,seth2019securing} \\
  
\hline

\multirow{4}{*}{\begin{tabular}{@{}c@{}c@{}c@{}@{}c@{}}   Public Health Informatics \\   \end{tabular}} & Infectious Disease Surveillance & \cite{hay2013big,young2014methods,garattini2019big} \\ 

 & Population Health Management & \cite{lamarche2014automated,cunha2015health,gamache2018public,van2019explainable,hatef2019public}  \\
 
 &  Mental Health Management &  \cite{nambisan2015social,seabrook2016social,conway2016social,mohr2013behavioral} \\
 
 & Chronic disease Management & \cite{bhardwaj2018impact,tu2015cardiovascular,kupersmith2007advancing} \\
 
\hline

\multirow{4}{*}{\begin{tabular}{@{}c@{}c@{}c@{}@{}c@{}}   Medical Signal Analytics \\   \end{tabular}} & Signal Acquisition  & \cite{mccullough2010effect,lee2010hypotensive,ahmad2009continuous,sun2010system,rajan2018internet} \\ 
 & Signal Storing from Monitoring Devices & \cite{adrian2013mongodb,kaur2015managing,han2006infrastructure,bressan2012trends,cao2008predicting,le2014consensus}  \\
 
 & Signal Integration and Aggregation  &  \cite{santos2011enabling,berndt2001healthcare,zhang2018sensor,nanda2015quantitative} \\
 
\hline

\end{tabular}
  \end{center}
\end{table*}

\section{Big Data Analystics Applications}

Healthcare sector produces huge amounts of patient data on a daily basis. Traditionally, most of this data was used to be in the form of hard copies but, due to the advancement in data acquisition devices, healthcare organizations are gathering data electronically. healthcare data analytics has the potential to bring in dramatic changes in healthcare industry to smooth the process and improving the quality of care. Data analytic researchers, healthcare providers, government agencies and the pharmaceutical companies identify range of different ways that big data techniques can help us to significantly improve patient outcome through policy making and evidenced based decisions. Below are the major areas in healthcare sector where big data analytics has a huge impact:

\textbf{Strategic Planning: }\textit{`Management is based on early measure: you cant manage if you cant measure'}. Healthcare is a time critical service. Hospitals  are struggling with patient flow. Machine learning and data analytics plays important role in the prediction of patient flow and ensuring smooth patient flow as well as reducing waiting period. Early predicting of hospital visit helps the management to decide and take the necessary step to reduce patient waiting time thereby giving timely treatment. Patient Flow Manager, Q-nomy's are the application that provides a comprehensive graphical view patient flow information, drawing of inpatient, elective, emergency, outpatient and other hospital systems. For example,  care mangers can analyze check-up results among patient in different demographic groups that help to identify what factors discourage patient from taking up treatment. The classical example is staff management: how many clinicians and nurcess staff should be give at specific time. 

For our first example of big data in healthcare, we will look at one classic problem that any shift manager faces: how many people do I put on staff at any given time period? If you put on too many workers, you run the risk of having unnecessary labor costs add up. Too few workers, you can have poor customer service outcomes – which can be fatal for patients in that industry. In other example, we can predict admission trend based on admission history of last few years i.e. using 10 years worth of hospital admissions records, which data scientists crunched using “time series analysis” techniques followed by machine learning  relevant to predicted future admissions trends.

\textbf{Fraud Detection:} \textit{`Suspect, detect and protect'}. Fraud, waste, and abuse have caused significant cost and it range from honest mistakes that result in erroneous billings, inefficiencies that may result in wasteful diagnostic tests, over-payments due to false claims. Personal data is extremely sensitive due to its profitable value in black-markets, thus, healthcare industry is 200\% more likely to experience data breaches than any other. With that in mind, effective detection of frauds is very important for reducing the cost and improving the quality of healthcare system. Fraud detection in healthcare is an important yet difficult problem. Big data has inherent security issues and healthcare organization are more vulnerable than they already are. Many organizations are using analytics to reduce security threats by analyzing the changes in network traffic, or suspicious behavior that reflects a cyber-attack. WhiteHatAI Centaur system, NICE ACTIMIZE, NHCAA, SAS, and Optum etc. are being used for medical claims processing that identifies and detects healthcare fraud, waste, and abuse before it happens. Likewise, data analytic can helps to prevent fraud and inaccurate claims in a systemic, repeatable way by streamlining the process of insurance claims. For example, the Centers for Medicare and Medicaid Services saved over $\$210.7$ million in frauds in just one year.

\textbf{Resource Management:} \textit{`How you use a facility, many factors pushing and pulling'}. Big data is making huge advances in reducing hospital waiting lists. Despite expensive efforts by the government and healthcare organizations, waiting times barely changed, with the median even increasing slightly i.e. Australia has been trying hard to reduce the waiting list times on its hospital for more than two decades. Efficient and timely  resource utilization helps to over come the patient flow and reduces the financial burden on organization. Data analytics continues to make inroads the manage hospital resources efficiently with respect to patient flow and risk. Examples are readmission, ambulance, and bed utilization etc.  

The common example is 30 days patient readmission or return visit to an emergency department. 30 days readmission identify the patients that have high possibility to return to hospital with 30 days of discharge. The development of risk prediction model helps to identify patients who would benefit from the disease management program in an effort to not only reduce the patient readmissions but also healthcare cost.

\textbf{Personalized Medicine:} \textit{`Disease and its treatment is unique as we are'}. The promise of personalized medicine is the shift away from ‘one size fits all’ medicine. Through the datafication and genomic fingerprints, much more information of each patient can be analyzed without requiring multiple rounds of testing. Best treatment can be made on an individual basis at a faster rate bu using personalized data.

\textbf{Genomics:} \textit{`The more is the data you have the better you can treat'}. Human body consists of 30,000 to 35,000 genes  \cite{drmanac2010human,international2001initial}. From the DNA structure of the human, it is estimated that there are 23 chromosomes with the distribution of 3.2 billion base pairs \cite{energy2011insights,hey2003data}. This data increases dramatically to about 200 gigabytes.  Thus big data analytics is required for genomics and sequencing practices that are used for the treatment of complex diseases like Crohn’s disease and age-related muscular degeneration \cite{koboldt2013next}. The impact of genomic data analytics has the great potential to improve healthcare outcomes, quality, and safety, as well as cost savings.

\textbf{Disease Prediction and Prevention: }\textit{`Precaution and care can help live longer' }. Many healthcare organizations, research labs, hospitals are leveraging Big Data analytics are by changing the models of treatment delivery. Thus Big Data analytics have tremendous applications in the healthcare domain for reducing cost overhead, detecting and curing diseases, predicting epidemics and enhancing the worth of human life by averting deaths. Number of projects from "Google", "DeepMind", "IBM", "Royal Free London NHS Foundation Trust" and "Imperial College Healthcare NHS Trust" and others have proved the importance of deep learning and machine learning for detection, identification, diagnostics and predictive analytics. DeepMind collaborated with Moorfields Eye Hospital to the analyze anonymized eye scans, searching for early signs of diseases leading to blindness. There are also projects signed with the Royal Free London NHS Foundation Trust and Imperial College Healthcare NHS Trust to develop new clinical mobile apps linked to EHR.

Big data has transformed healthcare by putting data to work, revealing clinical and operational insights. The most applicable applications of IBM are IBM Content and Predictive Analytics. "IBM Content and Predictive Analytics" for healthcare is the first industry-specific analytics solution to enable organizations to analyze the past, see the present and predict the future by simultaneously.  For example, we can predict admission trend based on admission history of last few years i.e. using 10 years worth of hospital admissions records, which data scientists crunched using “time series analysis” techniques followed by machine learning  relevant to predicted future admissions trends. One of the major application of big data analytics in the healthcare domain is medical image processing. As in healthcare enormous amount of medical images are produced like X-ray, CT and PET-CT images, MRI, ultrasound, fluoroscopy and photoacoustic imaging. These medical images produced big data that are used for various purposes like detection, diagnoses, assessment and decision making of therapy etc. \cite{ritter2011medical}.

Heart is the basic organ of the body. If the heart stops its working human body does not exist. There are several disorders of heart among them one is the heart attack. Big data analytics facilitates to predict the heart attack at the early stage using early heart attack detection system based on medical biosensor \cite{d2001validation,waqialla2016ontology} that detect heart attack at the early stage. There are some online systems \cite{alexander2017big}  and healthcare information system \cite{palaniappan2008intelligent} that provides guidance about heart diseases using IOT and Hadoop techniques.

The brain is the vital organ of the body that controls all the activities of the body just like CPU of the computer. Thus data mining and data analytics tools are deployed to detect the brain disorders like Parkinson's brain disease prediction  \cite{shamli2016parkinson,rehman2019deep}. Diabetics is one of the common diseases in this world. Big data analytics tools like \textit{`Hive'} and \textit{`R'} are used for the analysis of diabetics using descriptive dataset \cite{sadhana2014analysis,daghistani2015discovering}.  Efficient predictive models are established to reveal the data related to the investigation of diabetics.

There are online applications that are remotely facilitating the healthcare domain.  AmWell, Practo, Portea, and Isabel etc. are the most popular apps that are used for various purposes like appointment of doctors at hospitals, clinics etc., patient diagnosis, ordering medicines, consultation with the doctor remotely for treatment etc \cite{panda2017big}. Summering the applications of big data analytics in healthcare \cite{raghupathi2014big,helm2014use}, it is beneficial to identify and diagnose the patient accurately and precisely. It is used for the prediction and management of health risks and obesity to efficiently detect the level of frauds. It reduced the cost, variations, and elimination of duplicate care and improper claim submission.

\section{Challenges and Open Research Issues}
The healthcare sector suffers from multiple challenges, ranging from new disease outbreaks to preserving an optimal operational efficiency. To overcome these challenges, data mining and data analytics in the development of applications of healthcare have tremendous potential, however, success hinges on the availability of quality data but there is no magic recipe to successfully apply data analytics methods on any problem. Thus, the successful development of data analytics based applications depends on how data is stored, prepared and mined. However, chemical analytics poses a series of challenges when dealing with a enormous amount of complex data. These challenges involve data complexity, access to data, regulatory compliance, information security and efficient analytics methods, inter-operability, manageability, security, development, re-usability, and maturity.

\subsection{Multiple Source Information Management}
In healthcare data analytics, the main goal is to analyze the real world medical data to perform prediction or classification task. One of the biggest hurdles in development of such application depends upon on the data structure i.e. how medical data is spread across many sources, how data is stored, prepared and mined.  One of the worst example of lack of data sharing is: a woman who was was suffering from mental illness and substance abuse, visited variety of local hospitals more than 900 times in a  period of less than 3 years in Oakland, California, USA. It results in heavy cost, extensive use of hospital resources and more important, harder for womanto get good care.

Healthcare data is data correlations are leveraging in longitudinal records i.e. complex, heterogeneous, distributed and dynamic data i.e. in the US alone, healthcare data extended to 150 exabytes in 2011 and is expected to reach the zettabyte scale soon. Despite the rapid increase in EHR adoption, there are several challenges around making this information useful, readable and relevant to the physicians and patients who need it most. One of the key challenges in the healthcare industry is how to manage, store and exchange all of this data. Inter-operability is considered to be one of the solutions to this problem. There exists a poor inter-operability in EHRs that creates big data analytics challenging in healthcare. Integration of different data sources would require developing a new infrastructure where all data providers can collaborate each other to share. Another challenge is data privacy that limits the sharing of data by blocking out significant patient identification information such as MRN and SSN. Healthcare needs to catch up with other industries that have already moved from standard regression-based methods to more future-oriented like predictive analytics, machine learning, and graph analytics. Big data technologies like Data ingestion, data modeling, and data visualization are integrated with existing tools to provide a supported enterprise solution.  

Big data management is one of the hard tasks as there is a big cluster of data that is monitored and managed. Most patients visit multiple clinics to try to find a reason for their disease and medical solution for their illness. To overcome this issue, several management tools are integrated that is overwhelming and cost effective strategy. Proficiently handling large capacities of medical imaging data and extracting possibly useful information is another hard task. Hospitals have yet to achieve a level of inter-operability, and without it, it is almost impossible to improve patient care. The US Health Department is aiming for inter-operability between disparate EHRs by 2024. Medical stakeholders (physicians, administrators, patients etc.) believe that inter-operability will improve patient care, reduce medical errors and save costs. Imagine having the insight and opinions of hundreds of IVF/PGD patients to assist your decision before undergoing treatment rather than only relying on a physician’s recommendations. Due to the importance of data integration, healthcare organizations are turning to the implementation of inter-operability. To achieve a high level of inter-operability, HL7, HIPPA, HITECH and other health standardization bodies have demarcated several standards and guidelines to assist organizations to know whether they meet inter-operability and security standards. The Authorized Testing and Certifying Body (ATCB) provides a sovereign, third-party opinion on EHR. Two types of certification (CCHIT and ARRA) are used to evaluate the system. The review process comprises standardized test scripts and exchange tests of standardized data. Healthcare industry needs to catch up with other fields that have already progress from standardization.

\subsection{Security and Privacy and Confidentiality}

Every stakeholder in the health industry has a role to play in ensuring the security and privacy of patient information. It is a shared responsibility. Patient privacy and information security are fundamental components of a well-functioning healthcare system that helps to accomplish better health outcomes, healthier people, and smarter spending. For example, a patient may not disclose certain information or may ask a physician not to record his health information due to a lack of trust and the perception that this information might not be kept confidential. This attitude puts the patient at risk and deprives physicians and researchers of important information as well as putting the organization at risk in terms of clinical outcomes and operational efficiency analysis. To reap the benefits, providers and individuals must belief that patients’ health information is kept private and secure. On the other hand, providers are facing several challenges in ensuring that privacy and security issues are managed at a standard that meets the patients’ satisfaction i.e. efficient data analysis without providing access to precise data in specific patient records. Security and privacy in data analytics poses several challenges, especially when it draws information from multiple sources. 

The major goal in healthcare is not to protect the patient’s privacy rather it is to save lives. The HIPAA (Health Insurance Portability and Accountability Act) of 1996 comes to mind when privacy is debated in the health sector. It delivers legal rights to patients concerning their personally identifiable information and establishes responsibilities for healthcare providers to defend and restrict its use or disclosure. With the escalation in the amount of healthcare data, data analytics researchers envisage huge challenges in ensuring the anonymity of patient information to avoid its use or disclosure. Limiting data access, unfortunately reduces information content which might be very important. Moreover, real data is not static but grows larger and varies over time and none of the existing techniques result in any convenient content being released in this scenario.

\subsection{Advanced Analyzing Techniques}

Technological advancements (wearable devices, patient-centered care etc.) are transforming the entire healthcare industry. The nature of health data has progressed, and currently, EHRs have simplified the data acquisition process with the help of the latest technology, but unfortunately, they don’t have the ability to aggregate, transform, or perform analytics on it. Intelligence is restricted to retrospective reporting that is insufficient for data analysis. A plethora of algorithms, techniques, and tools are available for the examination of complex data. Traditional machine learning deploys statistical analysis based on a sample of a total dataset. The use of traditional machine learning methods for this data is not efficient and is computationally infeasible. The combination of the huge volume of healthcare data and computational power lets the analysts to focus on analytics techniques which are scaled up to accommodate the volume, velocity, and variety of complex data. During the last decade, there has been a melodramatic change in the size and complexity of data thus, several emerging data analysis techniques have been presented.

Healthcare needs to catch up with other industries that have already progressed from traditional methods to advance methods like predictive analytics, deep machine learning, and graph analytics. Innovative analytics techniques need to be developed to interrogate healthcare data and gain insight into hidden patterns, trends, and associations in the data. It deduces relationships without the need for a specific model and enables the machine to identify the patterns of interest in huge unstructured data. As one example, a deep learning algorithm that observed data from Wikipedia learned on its own that California and Texas are both states in the U.S. It does not have to be modeled to understand the conception of a country and state, and this is a gigantic difference between older machine learning and emerging deep learning methods.

\subsection{Data Quality}

 Gone are the days when healthcare data was small, structured and collected exclusively in electronic health records. Due to the tremendous advancements in IT, wearable technology and other body sensors, data has become quite large (moving to big data), unstructured (80\% of electronic health data is unstructured), non-standard as well as in a multimedia format. This variety in data makes it challenging and interesting for analysis. Currently, the quality of healthcare data is a cause of concern for four reasons, incompleteness (missing data), inconsistency (data mismatch between within same or various EHR sources), inaccuracy (non-standard, incorrect or imprecise data) and data fragmentation. Data quality involves a group of different techniques, these being data standardization, verification, validation, monitoring, profiling, and matching. The problem of poor data in the health industry has reached epidemic proportions and introduces several pernicious effects, particularly in relation to disease prevention. The problem with dirty data is mostly related to missing values, duplication, outliers and stale records.
 
Although real-time data monitors (especially in ICUs) are partially used in most hospitals, real-time data analytics is not in practice. Hospitals are moving to real-time data collection and in the near future, real-time data analytics will revolutionize the healthcare industry, enabling such things as the early identification of infections, the continuous monitoring of the progress of treatment, and the selection of the right drugs etc. which could help to reduce morbidity and mortality. To achieve real-time data processing, we need data standardization and device inter-operability.

The other common issue is data standardization. Structuring of only 20 percent of data has shown its importance but on the other hand, clinical notes are still in practice and created in billions due to the reason that the physician can best explain the clinical encounter. Empower physicians as well as maintaining the data quality is quite challenging. So far, this data is excluded from data analytics as it’s available in the natural language and not discrete. Transforming this unstructured data into a discreet form requires efficient intelligent technology and it is has been a very difficult problem for medical IT until now.  The only way this unstructured and nonstandard data can be used is by using NLP to translate the data using ICD or SNOMED CT into discrete data.

\section{Conclusion}

The exponential growth of big data analytics has rapidly increased that plays a vital role in the progression of healthcare practices and research. It includes providing tools to collect, analyze, manage and store a large volume of structured, unstructured and large complex data. Big Data has brought a dramatic change in healthcare which reduce the cost of treatment and accelerate the identification of disease, cancer etc. and improve the life's quality. It has been recently applied in aiding in the process of healthcare personnel, care delivery, early disease detection, disease exploration, patient care, and community services.

In this paper, we have discussed the big data analytics methods, tools, techniques and architectures in the healthcare domain. We have focused on five major sub-disciplines of healthcare i.e. medical image processing and imaging informatics, bioinformatics, clinical informatics, Public Health informatics and  Medical Signal analytics along with techniques, tools, and repositories deployed in each discipline. These disciplines plays a vital role in healthcare and bio medical due to the enormous amount of data. 

Healthcare providers had no direct incentive  in sharing the patient information with each other, that made it harder to efficiently utilize the power of analytics in healthcare industry. We can possibly change the way to healthcare providers use modern advances and sophisticated technologies to pick up understanding from their clinical, data warehouses, information storehouses for extracting informative patterns and decision making. Later on we'll see the quick, across the board execution and utilization of Big Data Analytics over the social insurance association and the medicinal services industry. Keeping that in mind, the few difficulties must be tended to. Its potential is extraordinary however, issues, for example, multiple source information management, ensuring protection, shielding security, setting up models and administration, advance analyzing techniques and data quality are the notable challenges in the domain. Regardless, the future trends of Big Data in the social insurance framework have the capability of enhancing and quickening communications among clinicians, executive, logistic manger, and analyst by diminishing costs, reducing risks and improving personalized care. 

Implementation of big data analytic is the responsibility for all stakeholders in healthcare industry. They must be effectively engaged in the review and policy making process if big data that could results in improving the patient outcomes. Government agencies, healthcare professionals, hardware companies, pharmaceutical industries, people, data scientist, researchers and vendors must be involved in developing the big data framework that will provide the future direction of big data analytics in healthcare industry.

%\begin{acknowledgements}
%\end{acknowledgements}
\scriptsize 
%\bibliographystyle{plain}
%\bibliography{bibl}

% \bibliographystyle{spmpsci}      % mathematics and physical sciences
% %\bibliographystyle{spphys}       % APS-like style for physics
% \bibliography{BigData}   % name your BibTeX data base
%\begin{thebibliography}{}
% etc
%\end{thebibliography}

\end{document}